%% file: ms_arxiv_accepted.tex
\documentclass[iop]{emulateapj}

\def\altomega{\varpi}
\usepackage{amsmath}
\usepackage{xcolor}

\input{preamble}

\slugcomment{Accepted to the ApJ.}

\makeatletter
\renewcommand\normalsize{\@setfontsize\normalsize\@xpt{12.5}}
\makeatother

\shorttitle{Dynamical Properties of MGCs in Galaxies at the EoR}
\shortauthors{Leung et al.}

\begin{document}
\title{Dynamical Properties of Molecular-forming Gas Clumps in Galaxies at the Epoch of Reionization}

\author{T. K. Daisy Leung\altaffilmark{1,2}}
\author{Andrea Pallottini\altaffilmark{3,4}}
\author{Andrea Ferrara\altaffilmark{4,5}}
\author{Mordecai-Mark Mac Low\altaffilmark{2,6}}

\affil{\textsuperscript{1} Department of Astronomy, Space Sciences Building, Cornell University, Ithaca, NY 14853, USA; }
\email{tleung@astro.cornell.edu}
\altaffiltext{2}{Center for Computational Astrophysics, Flatiron Institute, 162 Fifth Avenue, New York, NY 10010, USA}
\altaffiltext{3}{Centro Fermi, Museo Storico della Fisica e Centro Studi e Ricerche ``Enrico Fermi'', Piazza del Viminale 1, Roma, 00184, Italy}
\altaffiltext{4}{Scuola Normale Superiore, Piazza dei Cavalieri 7, I-56126 Pisa, Italy}
\altaffiltext{5}{Kavli Institute for the Physics and Mathematics of the Universe (WPI), University of Tokyo, Kashiwa 277-8583, Japan}
\altaffiltext{6}{American Museum of Natural History, 79th St.~at Central Park West, New York, NY 10024, USA}

\begin{abstract}
We study the properties of molecular-forming gas clumps (MGCs) at the epoch of reionization using cosmological zoom-in simulations. We identify MGCs in a $z\simeq 6$ prototypical galaxy (``\flower'') using an H$_2$ density-based clump finder.
We compare their mass, size, velocity dispersion, gas surface density, and virial parameter ($\alpha_{\rm vir}$) to \obs. In \flower, the typical MGC mass and size are $M_{\rm gas} \simeq 10^{6.5}$\,\Msun and $R\simeq45-100$\,pc, which are comparable to those found in nearby spirals and starburst galaxies. MGCs are highly supersonic and supported by turbulence, with r.m.s velocity dispersions of $\sigma_{\rm gas}\simeq$\,20\,--\,100\,\kms and pressure of $P/{\rm K}_B\simeq 10^{7.6}\,\rm K\,$\cc (i.e., $> 1000 \times$ with respect to the Milky Way), similar to those found in nearby and $z$\ssim2 gas-rich starburst galaxies.
In addition, we perform stability analysis to understand the origin and dynamical properties of MGCs. We find that MGCs are globally stable in the main disk of \flower. Densest regions where \SF is expected to take place in clouds and cores on even smaller scales instead  have lower $\alpha_{\rm vir}$ and Toomre-$Q$ values.
Detailed studies of the star-forming gas dynamics at the epoch of reionization thus require a spatial resolution of $\lesssim$40\,pc ($\simeq$\,0$\farcs$01), which is within reach with the Atacama Large (sub-)Millimeter Array 
and the Next Generation Very Large Array.
\end{abstract}
\keywords{methods: data analysis --
          galaxies: high-redshift --
          galaxies: ISM --
          galaxies: evolution --
          galaxies: formation --
          galaxies: starburst --
          stars: formation}


\section{Introduction}
The growth of galaxies and their subsequent evolution are governed by the baryon cycle---galaxies accrete gas from the intergalactic medium (IGM) either directly from the cosmic web, or through mergers with other galaxies. This gas fuels \SF and feeds central supermassive black holes. Subsequent feedback replenishes and enriches the circumgalactic medium by expelling some
part of this material.
Existing studies indicate that early galaxies are more gas-rich with molecular fractions higher than those of present-day galaxies \citep[e.g.,][]{vandevoort11b, Decarli16a, Decarli17a}. Massive gas inflows from the IGM trigger gravitational instabilities that lead to the formation of
dense gas clumps in which molecules form quickly (hereafter molecular-forming gas clumps\footnote{Note that by clumps,
we refer to kiloparsec-scale gas concentrations in galaxies, rather than molecular gas concentrations small than molecular clouds
and containing cores.};
MGCs) that are typically more massive ($M_{\rm gas}\simeq 10^9$\,\Msun) and extended ($\simeq$ sub-kpc) than those observed in nearby galaxies (e.g., \citealt{Gabor13a, Hopkins14a, Inoue16a}). Some theoretical works argue that the migration of such giant massive clumps through disks is largely responsible for the buildup of the bulges of massive galaxies at redshift \z$\sim$\,0 \citep[e.g.,][]{Ceverino10a}.

Early galaxies have higher star formation rates \citep[SFR; ][]{Behroozi13b, Sparre15a, Maiolino15a, Dunlop17a} and smaller sizes \citep[e.g.,][]{Bouwens11a, Ono13a} compared to the local population \citep[see also a review by][]{Stark16a}.
As a consequence, we expect them to be significantly more ionized, and have intense and hard interstellar radiation fields (ISRF). Since their metallicity and dust content are also expected to be lower in these early evolutionary stages, shielding of UV photons---responsible for the photoheating of the gas---  is strongly reduced. Such differences in turn affect the regulation of the thermal and chemical state of the multi-phase interstellar medium (ISM).

Studying ISM properties of early galaxies is essential for understanding how \SF proceeds under more extreme conditions.
Even in the local Universe, where detailed \obs can be made, variations in molecular cloud properties have been observed between different galaxy populations (see e.g., \citealt{Hughes10a, Hughes13b}).  Given that \highz galaxies statistically represent the early evolutionary stages of present-day galaxies, it is thus reasonable to pose the question: {\it what are the physical properties of
MGCs in early galaxies, and how do they differ from those found in local \galpop?}

FIR fine-structure lines (e.g., \cii, \nii, and \oiii), and CO and [\ci]~lines are key diagnostics for constraining the ISM conditions of galaxies. They also provide highly complementary information on different ISM phases (ionized, atomic, molecular; e.g., \citealt{Scoville74a, Rubin85a, Malhotra01a}).
Global measurements of these diagnostics in \highz galaxies have provided preliminary information on their global properties (e.g., gas masses, gas temperature, and radiation field intensity). However, spatially resolving their ISM is necessary to fully understand many aspects of galaxy evolution and the physics behind their intense \SF (SFR\ssim100$-$3000\,\Msun\,yr\pmOne; see e.g., the review by \citealt{CW13}).
To date, spatially resolved ISM properties of \highz galaxies have only been mapped in a handful of galaxies at high redshift using tracers such as dust continuum, CO, and \cii lines (e.g., \citealt{Swinbank11a, Hodge15a, Ferkinhoff15a, Hodge16a, Leung19a}). These studies find that galaxies close to the peak of cosmic \SF ($z$\ssim2) are more molecular gas-rich, turbulent, and clumpy than nearby galaxies, although luminous compact blue galaxies may represent local analogs of these gas-rich, clumpy systems \citep{Garland15}.

Earlier epochs still represent an essentially uncharted territory for ISM investigations. At present, it remains unclear how \SF proceeds in the (sub-)$L^*$ galaxy population at \z$\gtrsim$\,6  which is responsible for producing the bulk of the UV photons that reionized the Universe.
High resolution hydrodynamics simulations have been carried out to investigate the global properties, structures,
and morphologies of galaxies out to the epoch of reionization (EoR) and their importance for providing
the ionizing photons \citep[see e.g.,][]{Ceverino17a, Katz17a, Ma18a, Trebitsch18a, Rosdahl18a}.
Compared to the zoom-in galaxies presented in these works,
the simulation used here focuses on a Lyman-break galaxy \citep{Pallottini17a} from the \ncode{serra} simulation suite
(see \citealt{Pallottini19a} for a comparison between these zoom-in simulations).
Most notably, we include a non-equilibrium thermo-chemical network to follow the formation of H$_2$, which is
crucial to examining the properties of MGCs.

While ALMA has enabled the detection of \cii158\,$\micron$ and CO line emission in normal (SFR$<$\,100\,\Msun\,yr\pmOne) galaxies at \z$>$\,6 over the past few years \citep[e.g.,][]{Carniani18b, Odorico18a}, the first spatially resolved observations are just starting to become available \citep[e.g., ][]{Jones17a}.
On the theoretical side, a schematic investigation by \citet{Behrendt16} revealed that kiloparsec-scale clumps likely consist of many smaller clouds formed by gravitational instability, a hierarchical structure that persists in an idealized disk model by \citet{Behrendt19}.

To understand the physical properties of MGCs in early galaxies, we have undertaken a detailed numerical study whose aim is to characterize the dynamical properties of the star-forming MGCs in prototypical (i.e., $L^*$) galaxies in the EoR.

The paper is structured as follows\footnote{Throughout this paper, we adopt a concordance cosmology, with total matter, vacuum and baryonic densities in units of the critical density $\Omega_{\Lambda}$\eq0.692, $\Omega_m$\eq0.308, $\Omega_b$\eq0.0481, Hubble constant $H_0$\eq100\,$h$\,km s\pmOne\,Mpc\pmOne with $h$\eq0.678, spectral index $n$\eq0.967 and $\sigma_8$\eq0.826 \citep{Planck14a}.}. We start by providing some physical background in \Sec{Back}. In \Sec{sim}, we describe the setup of our simulation and properties of our main galaxy (\flower). In \Sec{eqn}, we describe the method used to identify MGCs, and present the formalism within which we interpret the results. In \Sec{results}, we present the results and characteristic properties of the MGCs. We then interpret the results and discuss the implications of our findings in \Sec{diss}, and give our conclusions in \Sec{conclusion}.

\section{Physical Background}\label{sec:Back}
We begin by introducing some empirical relations commonly mentioned in observational studies of molecular clouds in the literature, as well as describing the gravitational instability of galactic disks, which might be driving MGC formation. These concepts will be used in the subsequent analysis of our simulations.

\subsection{Larson's Relations}\label{sec:PVE}

\citet{Larson81a} discussed a number of relations among Galactic molecular cloud properties, namely the linewidth-size, density-size, and mass-size relations. Larson relations are routinely used for comparing properties of molecular structures in different galactic environments. They also represent a useful framework to analyze our results as they have been argued to arise from the interplay between gravity and velocity dispersion given by virial theorem (note, however, the alternative interpretation involving gravitational collapse by e.g., \citealt{Ballesteros-Paredes11a}).

The virial theorem for a distribution of unmagnetized gas can be written as \citep{McKee92a}
\begin{equation}\label{eqn:virial_th_general}
\frac{1}{2}\ddot{\mathcal{I}} = 2(\mathcal{T} - \mathcal{T}_{\rm ext}) + \mathcal{W},
\end{equation}
where $\ddot{\mathcal{I}}$ is the second time derivative of the
Lagrangian moment of inertia, $\mathcal{T}$ is the internal energy of
the gas (including thermal, turbulent, and bulk motions), $\mathcal{T}_{\rm ext}$ is external pressure support, and
$\mathcal{W}$ is the gravitational energy.
Let us specialize to the case of a spherical self-gravitating molecular cloud of mass $M_{\rm gas}$, radius $R$, and root-mean-square velocity dispersion $\sigma$, accounting for both thermal and turbulent contributions. Defining $P_{\rm ext}$ as the external pressure, \Eq{virial_th_general} can be written as
\begin{equation}
\frac{1}{2}{\ddot{\mathcal{I}}} = 3 M \sigma^2 - 4\pi P_{\rm ext} R^3 - \Gamma\frac{GM^2}{R}\,,
\label{eqn:virial}
\end{equation}
where $\Gamma$ is a geometrical factor that is equal to 3/5 for a uniform sphere; in \Eq{virial} the terms on the right-hand side represent the kinetic energy, external pressure, and gravitational potential energy terms.

Motivated by Larson's linewidth-size relation \citep{Larson81a} and the work by \citet{Heyer09a}, we assume equilibrium (i.e.,\ ${\ddot{\mathcal{I}}}=0$), define the cloud surface density as
$\Sigma$\eq$M/\pi R^2$, and rewrite the previous equation as
\begin{equation}
\frac{\sigma_{\rm gas}^2}{R} = \frac{1}{3}\left(\frac{4P_{\rm ext}}{\Sigma} + \frac{3}{5} \pi G \Sigma \right)\,,
\label{eqn:v0}
\end{equation}
which further reduces to
\begin{equation}
\frac{\sigma_{\rm gas}^2}{R} = \frac{\pi}{5} G \Sigma\,,
\label{eqn:SVE}
\end{equation}
if the external pressure $P_{\rm ext}=0$. For this case (often referred to as simple virial equilibrium) from the balance between kinetic and gravity terms we can define the virial parameter as
\begin{equation}
\alpha_{\rm vir} \equiv  \frac{5\sigma_{\rm gas}^2R}{GM_{\rm gas}} = \frac{5\sigma_{\rm gas}^2}{\pi G \Sigma R}\,,
\label{eqn:alpha}
\end{equation}

Based on Equation~\ref{eqn:SVE}, a one-to-one mapping between $\sigma_{\rm gas}^2/R$ and $\Sigma$ is therefore expected for a virialized cloud, since $\sigma_{\rm gas}^2/R\propto\Sigma$.
\citep{Heyer09a} pointed out that the original size-linewidth relation implies constant surface density in this interpretation. The limited dynamic range in column density of CO observations appears to account for this \citep[see also][]{Ballesteros-Paredes11a}.
Deviations from this relation are often attributed to a significant contribution from external pressure as per Equation~\ref{eqn:v0} (see e.g., \citealt{Heyer09a, Hughes10a, Hughes13b, Meidt13a}). Similar conclusions are also reached from the
analysis of clouds forming in a Milky-Way like galaxy simulated with a $\simeq 4\,{\rm pc}$ resolution \citep[i.e.,][]{grisdale:2018}.

Summarizing, the virial parameter can be used to quantify the stability/boundedness of a molecular cloud. Accounting for the external pressure, a virial parameter of $\alpha_{\rm vir}\lesssim2$ would be unstable \citep{Bertoldi92b}. Such a criterion is often used in \obs \citep[see e.g., ][]{Kauffmann17b}.

In cases where a stellar component plays an important role in the dynamics, the virial parameter becomes
\begin{equation}
\alpha_{\rm vir, tot} \equiv \frac{5 R}{G (M_{\rm gas} + M_{\star})} \frac{M_{\rm gas} \sigma_{\rm gas}^2 + M_{\star} \sigma_\star^2}
					       {M_{\rm gas} + M_{\star} }\,,
\label{eqn:alpha_tot}
\end{equation}
where $M_\star$ is the stellar mass enclosed within the MGC volume.

\subsection{Toomre Stability}\label{sec:Q}

The onset of gravitational instability is tightly connected to \SF
\citep[e.g.,][]{Kennicutt89a, Martin01, Wang94a, Li05b, Li06a}. For axisymmetric modes, the dispersion relation for the growth of density perturbations in a rotating, turbulent disk of finite thickness $h$ is described by
\begin{equation}
\omega^2 = \kappa^2 - \frac{2\pi G \Sigma |k|}{1 + |k| h} + \sigma_{\rm disk}^2 k^2\,,
\label{eqn:3Ddisp}
\end{equation}
where $k$ is the wavenumber and $\kappa$ is the epicyclic frequency, defined as:
\begin{equation}
\kappa^2\equiv\frac{2\Omega}{\altomega}\frac{d}{d\altomega}\left(\altomega^2\Omega\right)\,.
\label{eqn:kappa}
\end{equation}
\citep{Romeo92a},
where $\altomega$ is the galactic radius.
In Equation~\ref{eqn:3Ddisp} the terms on the right hand side are related to rotation, self-gravity and internal pressure, respectively. Heuristically, the instability can be understood by considering the scale at which gravitational potential overcomes the internal energy. Gravity dominates at scales $L > L_J$, where $L_J$ is the Jeans length. However, differential rotation in disk galaxies can stabilize perturbations that might otherwise collapse for $L > L_{\rm rot}$, where $L_{\rm rot}$ is set by $\kappa$. As a result, disks are unstable to gravitational collapse on scales between $L_J < L < L_{\rm rot}$.

From the dispersion relation, a parameter $Q$ can be derived such that $Q < 1$ when instability occurs, that reproduces this inequality to order unity. For a collisionless fluid---such as an ensemble of stars--- this parameter is \citep{Toomre64a}
\begin{equation}
Q_{\star} \equiv\frac{\sigma_{\star}\kappa}{3.36 G \Sigma_{\star}}\,.
\end{equation}
The equivalent parameter for a collisional gas was derived by \citet{Goldreich65a}
\begin{equation}
Q_{\rm gas}\equiv\frac{\sigma_{\rm gas}\kappa}{\pi G \Sigma_{\rm gas}}\,.
\label{eqn:Q}
\end{equation}
In the thin disk approximation ($kh\ll1$), instability occurs on scales $k$ such that $Q < Q_{\rm crit}\simeq1$ (or equivalently $\omega^2 < 0$ in \Eq{3Ddisp}). A frequently used observable proxy for $Q$ is the ratio of disk circular velocity to root-mean-square velocity dispersion $v_{\rm circ}/\sigma_{\rm disk}$ \citep[e.g.,][]{GarciaBurillo03a, Genzel11a, Kassin12a, Leung19a}.

In our stability analysis, we account for the combined effect of gas and stars \citep[derived exactly by][]{Rafikov01a}, and for the non-negligible disk thickness. This is done by adopting an approximation for an effective two-component $Q_{\rm eff}$ parameter \citep[i.e.,][see also \citealt{Inoue16a}]{Romeo11a, Romeo13a}.
The effect of disk thickness modifies the $Q$ parameter for gas and stars by accounting for the vertical velocity dispersion
\begin{equation}
T_{x} = \left\{
		\begin{array}{lccr}
			{\displaystyle 0.8 + 0.7\left(\frac{\sigma_{z}}{\sigma_{r}}\right)}      && & \mbox{if\ } \sigma_z \gtrsim 0.5 \times \sigma_r \\ [1.25em]
			{\displaystyle 1 + 0.6\left(\frac{\sigma_{z}}{\sigma_{r}}\right)^2}        & & & \mbox{if\ } \sigma_z \lesssim 0.5 \times \sigma_r \\
		\end{array}
	\right.
\end{equation}
and
\begin{equation}
Q^{\rm thick}_{x} = T_{x} Q\,,
\end{equation}
with $x$ indicating either gas or stars. The combined effect of gas and stars can then be accounted for by writing
\begin{equation}\label{eqn:q_eff}
Q^{-1}_{\rm eff} =  \left\{
				\begin{array}{lccr}
					     {\displaystyle\frac{w}{Q^{\rm thick}_{\star}} + \frac{1}{Q^{\rm thick}_{\rm gas}}}      & & & \mbox{if\ }  Q^{\rm thick}_{\star} \geq Q^{\rm thick}_{\rm gas} \\ [0.75em]
                                               {\displaystyle\frac{1}{Q^{\rm thick}_{\star}} + \frac{w}{Q^{\rm thick}_{\rm gas}}}      & & & \mbox{if\ } Q^{\rm thick}_{\star} \leq Q^{\rm thick}_{\rm gas} \\
				\end{array}
			    \right.
\end{equation}
where the relative weight $w$ is defined as
\begin{equation}
w\equiv\frac{2 \sigma_{\star} \sigma_{\rm gas}}{\sigma_{\star}^2 + \sigma_{\rm gas}^2},
\end{equation}
Conceptually, the finite disk thickness reduces the gravity in the vertical direction, thereby making it easier for a system to maintain stability, and thus lowering the critical Toomre $Q_{\rm crit}$ from $\simeq$\,1 to 0.67 \citep{Goldreich65a}.
On the other hand, including the contribution of the stellar component promotes gravitational instability, and thus increases
$Q_{\rm crit}$, more-so if the stars have low velocity dispersion.
As a rule of thumb, $Q_{\rm crit}\eq1.34$ for $Q_{\rm gas}$\eq$Q_\star$.  

\section{Numerical Simulations}\label{sec:sim}
The simulations used in this work are described by \citet{Pallottini17a, Pallottini17b} and are briefly summarized here.
\ncode{Serra}\footnote{Greenhouse in Italian.} is a suite of cosmological zoom-in simulations performed using Eulerian hydrodynamics and adaptive mesh refinement (AMR) techniques to achieve high spatial resolution in regions of interest (e.g., regions of high density).
In particular, it uses a modified version of \ncode{ramses} \citep{Teyssier02a} as the AMR backend. The simulation used here covers a comoving box of 20\,Mpc $h$\pmOne in size. The simulation zooms in on a target halo of mass $M_{\rm DM}\simeq10^{11} \Msun$
at $z$\eq6. The Lagrangian region of the halo (2.1\,Mpc $h$\pmOne) has a dark matter mass resolution of $\simeq 6\times 10^4 \Msun$, equivalent at initial density to a baryonic mass resolution of $1.2 \times 10^4 \Msun$. This region is spatially refined with a quasi-Lagrangian criterion based on a mass threshold, so that a cell is refined if its total (dark+baryonic) mass exceeds the mass resolution by a factor of eight. The finest refined cell allowed in the zoom in region has size $l_{\rm cell}\simeq$\,30\,pc (at $z = 6$), i.e., sizes are comparable to the sizes of local giant molecular clouds \citep[e.g.,][]{Sanders85a, Federrath13a, Goodman14a},

The models include a non-equilibrium chemical network (\ncode{Krome}) following e$^{-}$, H$^+$, H$^-$, He, He$^+$, He$^{++}$, H$_2$,
and H$_2^+$ \citep{Grassi14a,Bovino16a}. Of particular importance here is that the abundances are calculated using an on-the-fly non-equilibrium
formation of molecular hydrogen scheme described by \citet{Pallottini17a}.
In the simulation, the UV radiation field is assumed to be uniform spatially.
Self-shielding of H$_2$ from photo-dissociation is accounted for using the \citet{Richings14a} prescription,
while the formation rate on dust grain is computed following \citet{Jura75a}. Photo-dissociation via Lyman-Werner photons is included as
part of the \ncode{Krome} network. 
We do not assume a clumping factor in the chemistry solver (i.e., $C_\rho$\eq1, cf. \citealt[][]{Lupi20a}).
In the simulation, \SF follows an H$_2$-based Schmidt-Kennicutt relation.

Star formation is modeled using an H$_2$-based prescription of the Schmidt-Kennicutt relation \citep{Krumholz09a}. We adopt stellar tracks from \ncode{starburst99} and include stellar feedback from supernovae (SNe) as well as winds from OB and asymptotic giant branch stars. Coupling to the gas is implemented via a sub-grid model for blastwaves, which accounts for radiative energy losses inside the cell.
The remaining energy is injected into the ISM in both thermal and non-thermal form \citep{Agertz13a}; the latter is a term that mimics unresolved turbulence. The non-thermal energy is not affected by radiative cooling but rather is dissipated on the eddy turn-over time scale \citep{Maclow99a}. See \citet{Pallottini17b} for further details.
Radiation pressure on dust and gas is also included. Photoionization from local sources is neglected in this simulation \citep[see][for its effect]{Pallottini19a, Decataldo19a}.

The simulation zooms in on a galaxy named after the flower \flower, whose properties are given in \citet{Pallottini17a} and are briefly summarized in the following. \flower is a Lyman-break galaxy that at $z\simeq 6$ is hosted by a dark matter halo of mass
$M_{\rm DM}\simeq$\eq3.5\,$\times$\,10$^{11}$\,\Msun at the center of a cosmic web knot, and accretes mass from the IGM mainly via three filaments of length $\simeq$\,100\,kpc. \flower has a stellar mass of $M_\star\simeq$\,2.6\E{10}\,\Msun, a metallicity of $Z\simeq$\,0.5\,$Z_{\odot}$, a molecular gas mass of $M_{\rm H2}\simeq$\,5\E{7}\,\Msun, and a SFR of 30\,--\,80\,\Msun\,yr\pmOne.
The specific SFR of \flower is ${\rm sSFR} \equiv {\rm SFR}/M_{\star} \simeq$~4--40 ~Gyr\pmOne, compatible with the sSFR vs.\ $M_{\star}$ relation observed at high redshift \citep{Jiang16a}. The effective stellar radius of \flower is $\simeq$\,0.5\,kpc and the dark matter virial radius is $r_{\rm 200}\simeq$\,15\,kpc.
The stellar-to-halo mass ratio of \flower is $M_\star/M_{\rm DM}\,\simeq\,$0.07.
This is a factor of about three higher than the typical value inferred by abundance matching models, which typically give values of $\sim0.02$.
But studies show that this ratio varies as the halo mass becomes much larger or smaller than from $M_h$\ssim10$^{12}$\,\Msun
(see \citealt{Moster13a, Behroozi13b, Behroozi15a, Moster18a})\footnote{ 
This ratio from abundance matching can vary; for instance, \citet{Behroozi13b} (main method) finds a ratio of 0.015\,--\,0.025
for a halo mass between $M_{\rm halo}$\ssim10$^{10.5}$--$10^{12}$\,\Msun, whereas in the same paper the authors find
0.01--0.05 in the same mass range from direct abundance matching
(see the Appendix therein).
As another example, from an earlier abundance matching study, \citet{Moster13a} find a ratio 0.005, also differing by a factor three from \citet{Behroozi13b}, though in opposite direction.}.
The ratio of \flower is higher than e.g., \citet{Katz17a}  and \citet{Ceverino17a} which could be a result of different feedback implementation.

\subsection{Star Formation History} \label{sec:sfh}

\begin{figure*}[hptb]
\centering
\includegraphics[trim=0 0 0 0, clip, width=0.65\textwidth]{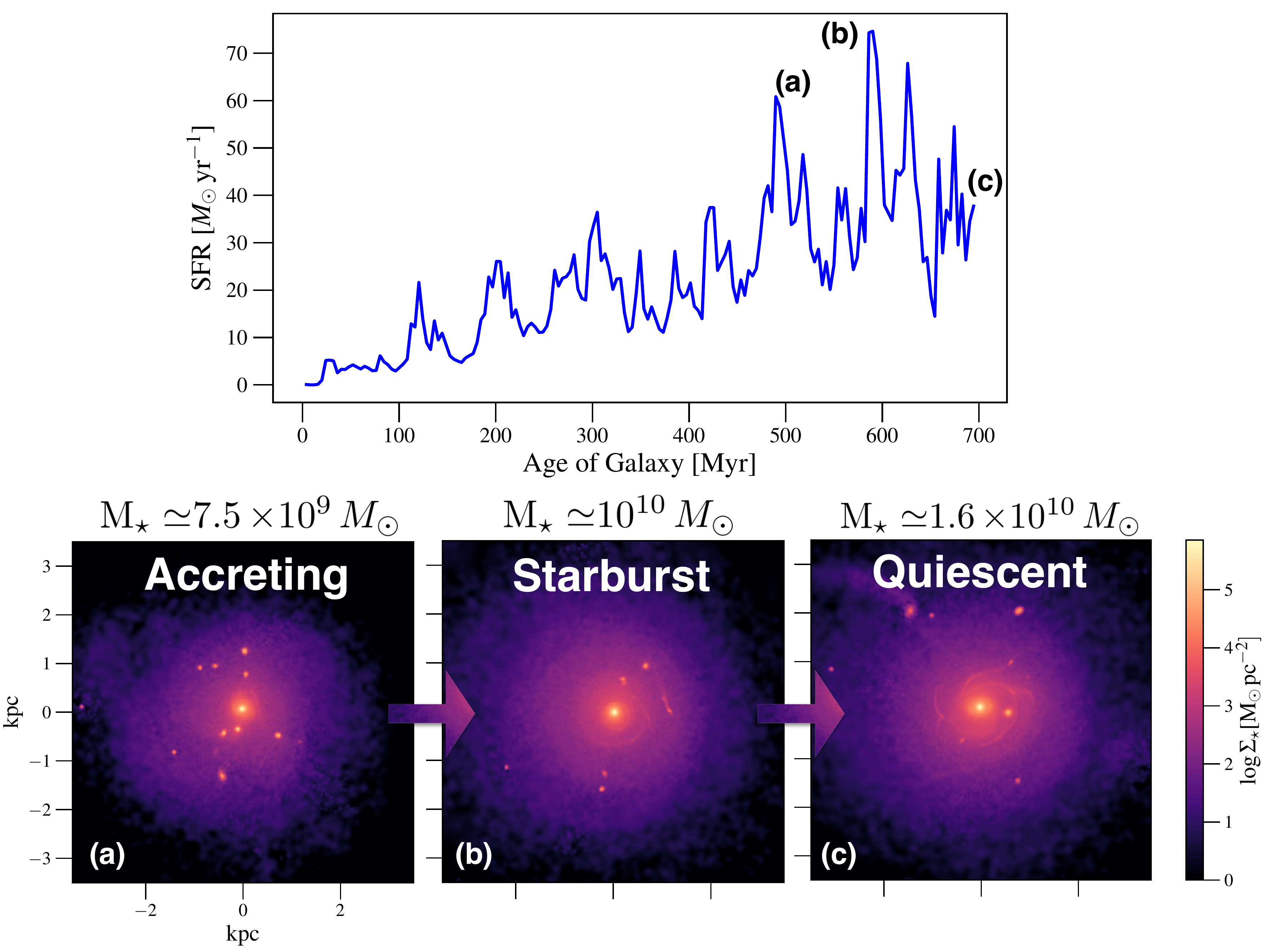}
\caption{
    {\it Top}: Star formation history of \flower. {\it Bottom}: projected stellar mass distribution during {\it (a)} an early accreting phase;
    {\it (b)} a major starburst following a merger event; and {\it (c)} a relatively quiescent post-starburst phase.
    \label{fig:SFH}
   }
\end{figure*}

One of the main advantages of studying galaxies in simulations is that we can examine how their dynamical properties evolve with time, which is interesting in order to understand the physical processes that determine the morphology and dynamical properties of galaxies, which in turn affect their \SF.
This is advantageous especially at early cosmic epochs, when the densest structures are beginning to form; gas is constantly being accreted onto the central galaxy from the cosmic web and satellite galaxies, thereby leading to bursts of \SF.
Meanwhile, tidal forces resulting from interactions with these surrounding galaxies can disrupt the main disk and arms, likely leading to different dynamical states for the molecular structures compared to more evolved galaxies found at a later cosmic time (e.g., some molecular structures may disperse while others may agglomerate into more massive ones).
The \SF history\footnote{Note that here the SFR is calculated based on the stellar mass formed in the last 4 Myr within 3.5 kpc from the galaxy center of mass. The SFR plotted in Figure 2 of \citet{Pallottini17b} is a factor of two higher since there the SFR accounts for the contribution from massive satellite galaxies within the virial radius ($\approx$15\,kpc).} of \flower is shown in \Fig{SFH}. The SFR of \flower varies between $\sim$30--80 \Msun\,yr\pmOne as it evolves from an actively accreting phase to a starburst phase after a merger, and then back to a relatively quiescent phase, over the simulated $\approx 700$\,Myr.

In \Sec{dist}, we show the importance of rotation support from large-scale motions in the MGC dynamics by comparing their properties in the disturbed phase of \flower and in the disk-like ordered phase.
Given the stochastic nature of \flower in its star formation history, we mostly focus on few of its most extreme evolutionary stages in this work (see \Sec{singless}).
These phases correspond to (a) an intensely accreting phase and (b) a starburst phase (\Fig{SFH}). We are interested in determining whether MGC properties are sensitive to these different dynamical conditions. For completeness, we also show relationships of MGC properties examined for other evolutionary stages traced in the simulation (see \Sec{ncut}).

\section{Molecular-forming Gas Clumps}\label{sec:eqn}

\subsection{Identification}\label{sec:method}

To identify the molecular complexes, we use a customized version of the clump-finding algorithm available in the \ncode{python} package \ncode{yt} \citep{Turk11a}, which was initially described in \citet{Smith09a} and modified since.
The latest version of the default \ncode{yt} clump finder decomposes the zones of the simulation into non-overlapping tiles, which are stored in a three-dimensional tree that can be processed using $k$-dimensional tree algorithms. It then identifies the contours of a variable field (here, the density field) within a tile and connects them across the tiles. In the customized version used for this study, we enhanced the stability of the code.
Due to the nature of our AMR simulation, we regrid the simulation data into uniform grids. The grid size is defined based on the highest resolution of the simulation data, i.e., the less refined regions are supersampled in the resulting uniform grids.

In the clump-finding process (in position-position-position space; PPP), we employ a set of different density thresholds defined based on the molecular hydrogen density of \flower at different evolutionary stages ($z=6.0$--7.2).
We note that this process is the three-dimensional analog to identifying molecular structures based on the noise levels in position-position-velocity (PPV) maps that observers obtain with telescopes, using molecular line tracers such as CO, CS, and HCN. This is commonly done by identifying clumps based on/after applying S/N-clipping, using tools such as the \ncode{aips} tasks \ncode{serch}, \ncode{clumpfind}, and \ncode{cprops}; (e.g., \citealt{Williams94a, Oka01a, Rosolowsky06a, Rosolowsky08a, DonovanMeyer13a})).
Existing studies find a good correspondence in the dynamical properties extracted in PPV- versus PPP-space for well-isolated structures (\citealt{Ballesteros-Paredes02a, Heitsch09a, Shetty10a, Beaumont13a, Pan15a}, but see \citealt{Ballesteros-Paredes02a} and \citealt{Shetty10a} for a discussion on caveats and limitations).

\begin{figure}[htbp]
\centering
\includegraphics[trim=0 0 0 0, clip, width=0.5\textwidth]{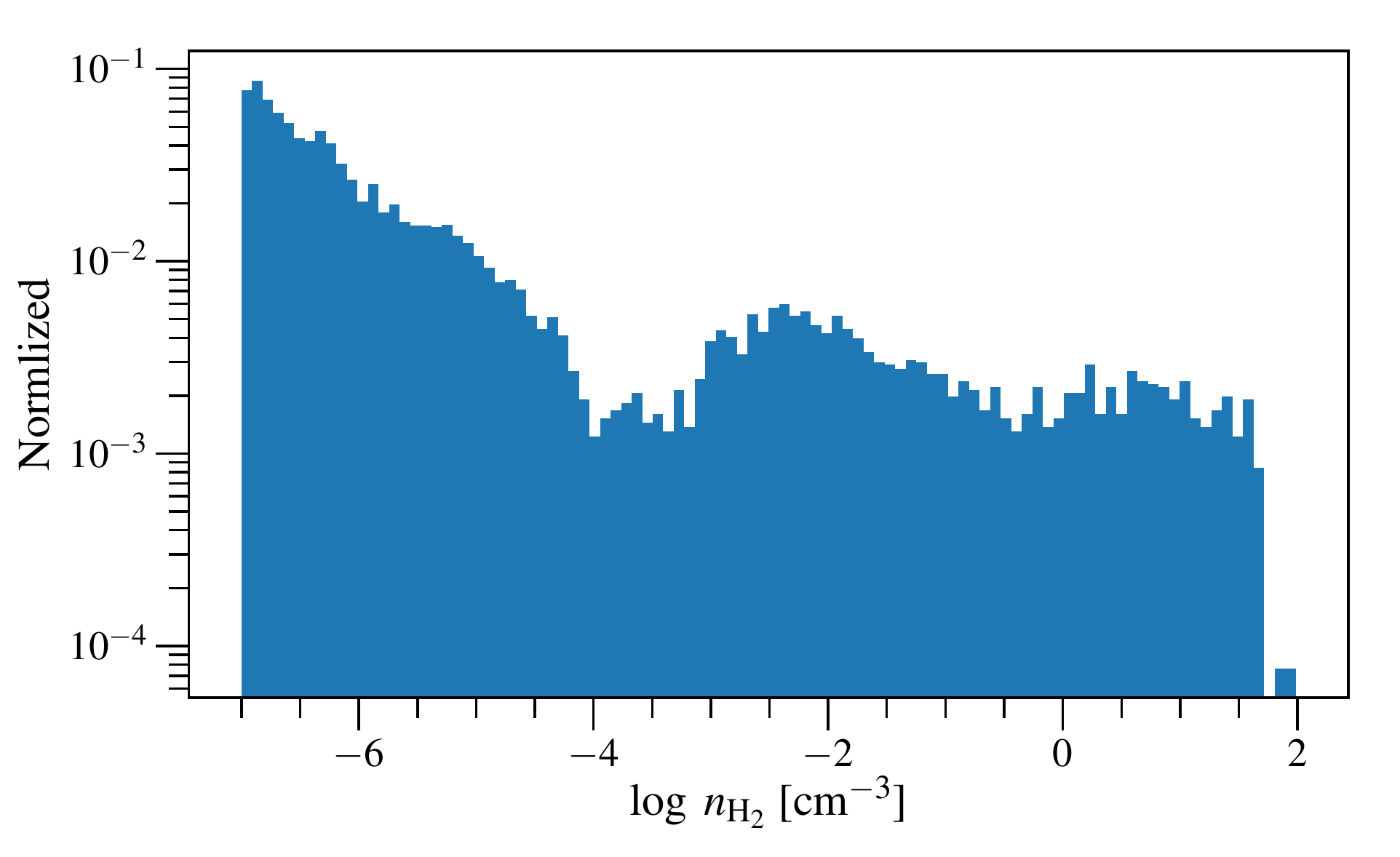}
\caption{Normalized distribution of molecular gas density of \flower during the accreting phase shown in \Fig{SFH}{\em
    (a)}.
\label{fig:h2density}}
\end{figure}

\begin{figure*}[htbp]
 \centering
  \includegraphics[width=\textwidth, trim=25 25 35 30, clip]{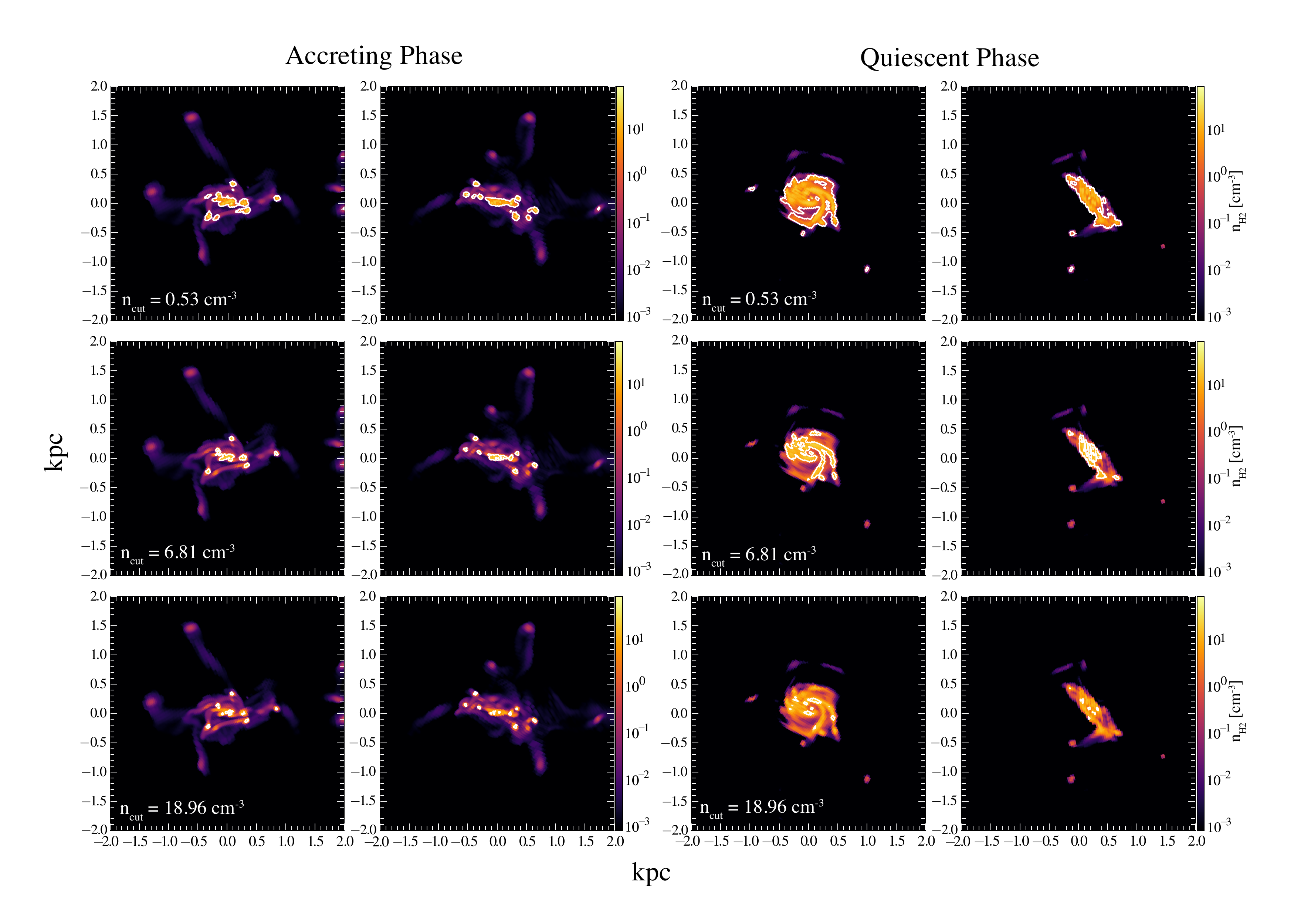}
\caption{
Examples of MGCs identified in \flower by our clump finder.
The identified structures (white contour) are superimposed on the map of the mass weighted $H_2$ density field of the galaxy.
Two stages of \flower are selected: the accreting phase (left two columns, see \Fig{SFH}a for the definition) and the quiescent phase (right two columns), where \flower display a highly disturbed gas morphology, and a rotating disk configuration, respectively.
Panels in the second and fourth columns show the molecular gas distribution projected along different viewing angles.
Different rows show the results obtained by applying different H$_2$ number density cuts $(n_{\rm cut})$, as annotated in each panel.
\label{fig:MGC}
}
\end{figure*}

In \Fig{h2density}, we show the distribution of H$_2$ number density $n_{\rm H2}$ for \flower during its accreting phase, including the contribution from the gas within 3.5 kpc from the galaxy center.
We note that the distribution is almost flat for $n_{\rm H2}\gtrsim1$\,\cc and it samples the range of densities where clumps are found based on morphological analysis\footnote{For $n_{\rm H2}\gtrsim1$\,\cc, the fourth Minkowski functional of the H$_{2}$ density field is significantly larger than zero. This implies that the field is made of isolated components. See Fig. 6 in \citet{Pallottini17b}.} \citep{Pallottini17b}.
In each evolutionary stage, we identify MGCs by applying density cuts to the {\em H$_2$ density} distribution, i.e., MGCs are selected at $n_{\rm cut} \leq n_{\rm H2}$. We select 10 equally-spaced cuts in H$_2$ density in log scale: ${(n_{\rm cut}/{1 {\rm cm}^{-3})}}\eq[0.32, 0.53, 0.88, 1.45, 2.45, 4.08, 6.81, 11.36, 18.96, 31.62]$. Note that with these choices we are including MGCs that are not fully molecular ($n = 0.5 n_{\rm H2}$).

To visually display the clump finding procedure, we overplot the molecular structures identified using a subset of the H$_2$ density cuts ($n_{\rm cut}$\eq0.53, 5.81, and 18.96\,\cc) on the H$_2$ density maps (\Fig{MGC}). Since the molecular structures are identified in the 3D H$_2$ density field, they can appear as overlapping structures depending on the viewing angle; thus we also plot them in different projections so that the identification can be more easily appreciated.
We repeat this identification process for 14 evolutionary stages between redshift \z$\in$[6.0, 7.2], spaced by $\Delta t$\eq15\,Myr.

We impose the additional constraint that an identified structure must be composed of at least 10 cells. We caution that an important caveat of such a constraint is that we can only examine the parameter space of cloud complexes of radius $R\gtrsim 40$\,pc, because of the resolution limit of the simulation.

\subsection{MGC Properties} \label{sec:distribution}

Upon identifying the molecular structures, we extract properties such as the gas mass $M_{\rm gas}$, effective size $R$, Mach number
$\mathcal{M}$, velocity dispersion $\sigma_{\rm gas}$, and gas surface density $\Sigma_{\rm gas}$ to examine their dynamics.

The mass of an MGC is calculated from the uniformly-gridded 3D density field, integrating over the MGC volume $V$. The effective size is defined assuming spherical geometry, i.e., $R \equiv (3 V /4 \pi)^{1/3}$.
The full velocity dispersion of MGCs is calculated from the bulk velocity field ($\sigma_{\rm bulk}$), thermal sound speed ($c_s$), and the non-thermal velocity dispersion ($\sigma_{\rm NT}$)
\begin{equation}
\sigma_{\rm gas}^2 = \sigma_{\rm bulk}^2 + c_s^2 + \sigma_{\rm NT}^2.
\label{eqn:veldisp}
\end{equation}
In \obs of MGCs, the linewidth contribution of dense gas exceeds that from the diffuse gas. Therefore, when calculating the global quantities of MGCs, we always perform a mass-weighting. Since we operate on data on a uni-grid, this is equivalent to density averaged quantities. In general, for the quantity $x$ in a MGC we write
\begin{equation}\label{eqn:defineaverage}
\langle x \rangle \equiv \frac{\sum_{i} \rho_i x_i }{\sum_i \rho_i}\,,
\end{equation}
where the sum is done for the cells indexed by $i$ composing the MGC.
We use the definition \Eq{defineaverage} to write each term of the right-hand side of Equation~\ref{eqn:veldisp} as follows.
%
The bulk velocity dispersion is
\begin{equation}
\sigma_{\rm bulk}^2 = \frac{1}{{3}} \langle \left|\mathbf{v} - \langle \mathbf{v} \rangle  \right|^2\rangle \,,
\end{equation}
where \Eq{defineaverage} is applied to each velocity component.
The thermal sound speed is calculated from the thermal pressure ($P_{\rm TH}$) through
\begin{equation}
c_s^2 = \left\langle \frac{{\rm k_B} ~P_{\rm TH}}{{\rm m_p}\, n} \right\rangle\,,
\end{equation}
where the pressure is in units of K\,\cc, $\rm k_B$ is the Boltzmann constant, $m_p$ is the mass of a proton. Similarly, contribution from non-thermal energy is calculated from non-thermal pressure ($P_{\rm NT}$) as follows
\begin{equation}
\sigma_{\rm NT}^2 = \left\langle \frac{{\rm k_B} ~P_{\rm NT}}{{\rm m_p}\, n}\right\rangle\,.
\end{equation}
Finally, the Mach number is related to the pressure terms as follows
\begin{equation}
\mathcal{M} = \left\langle \sqrt{ 1 + \frac{P_{\rm NT}}{P_{\rm TH}}} \right\rangle.
\label{eqn:MPress}
\end{equation}

\section{Results}\label{sec:results}

\subsection{MGC Basic Properties} \label{sec:dist}
\begin{figure*}[htbp]
\centering
\includegraphics[trim=0 0 0 0, clip, width=\textwidth]{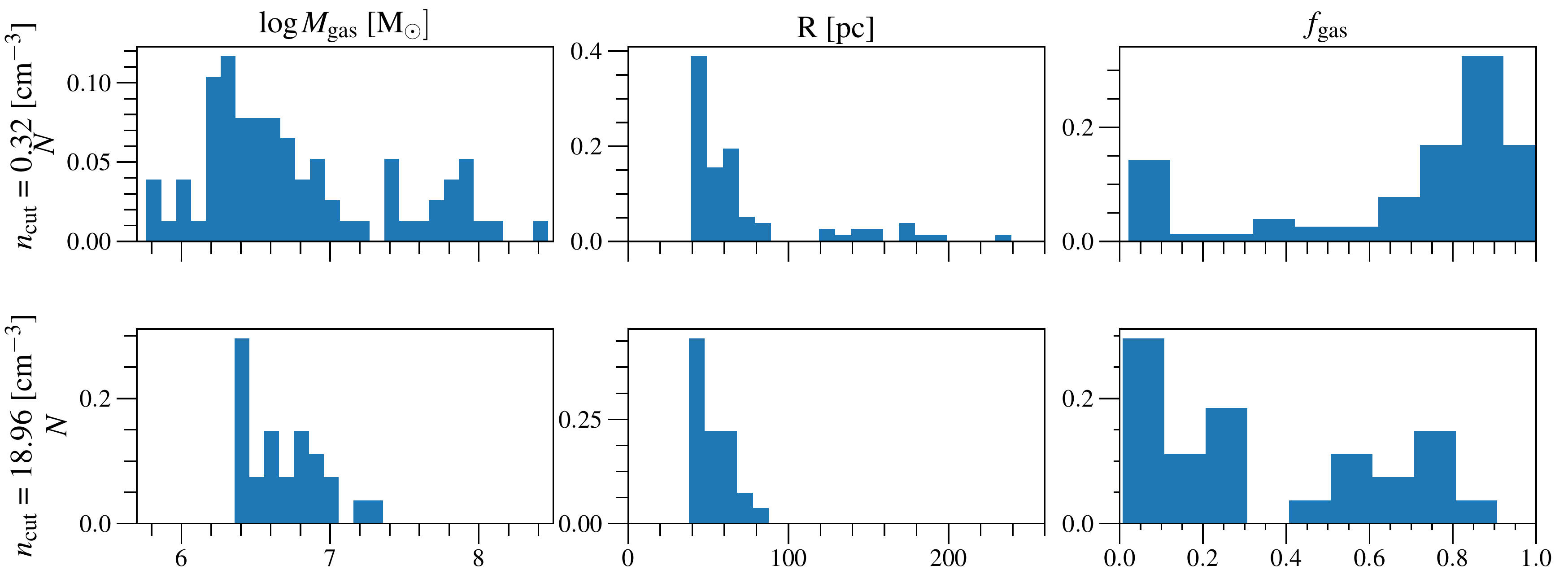}
\caption{Normalized distributions of mass (left), size (middle), and gas mass fraction (right) of MGCs identified using the lowest $n_{\rm cut}$\eq0.32\,\cc (top panels) and $n_{\rm ncut}$\eq18.96\,\cc (bottom panels) over all the considered evolutionary stages of \flower traced in the simulation. Note that the scales shown on the $y$-axes are different between the top and bottom panels, as fewer MGCs are identified at higher $n_{\rm cut}$.
\label{fig:dist}}
\end{figure*}

We start by considering the distributions of MGC total gas mass $M_{\rm gas}$, radius $R$, and gas mass fraction
\begin{equation}
f_{\rm gas} = \frac{M_{\rm gas}} {\left(M_{\rm gas} + M_\star\right)}.
\end{equation}
We show in \Fig{dist} the distributions from the combination of the three considered evolution stages. We show results for a low density cut of $n_{\rm cut}$\eq0.32\,\cc (top panels) and a high cut\footnote{The highest density threshold we used of $n_{\rm cut}$\eq31.62\,\cc corresponds to a minimum MGC mass of the order of 10$^{5.5}$\,\Msun for the densest structure. However, not all evolutionary stages considered have at least one MGC at this highest density threshold, so we consider MGCs identified with the second highest density threshold instead.} of $n_{\rm cut}$\eq18.96\,\cc (lower panels).
Overall, the mass of all MGCs identified ranges from $M_{\rm gas}\simeq$10$^{5.5-8.5}$\,\Msun,
whereas the gas fraction ranges from $f_{\rm gas} \simeq$\,0.1\,--\,1.

\begin{figure}
\centering
\includegraphics[width=0.5\textwidth]{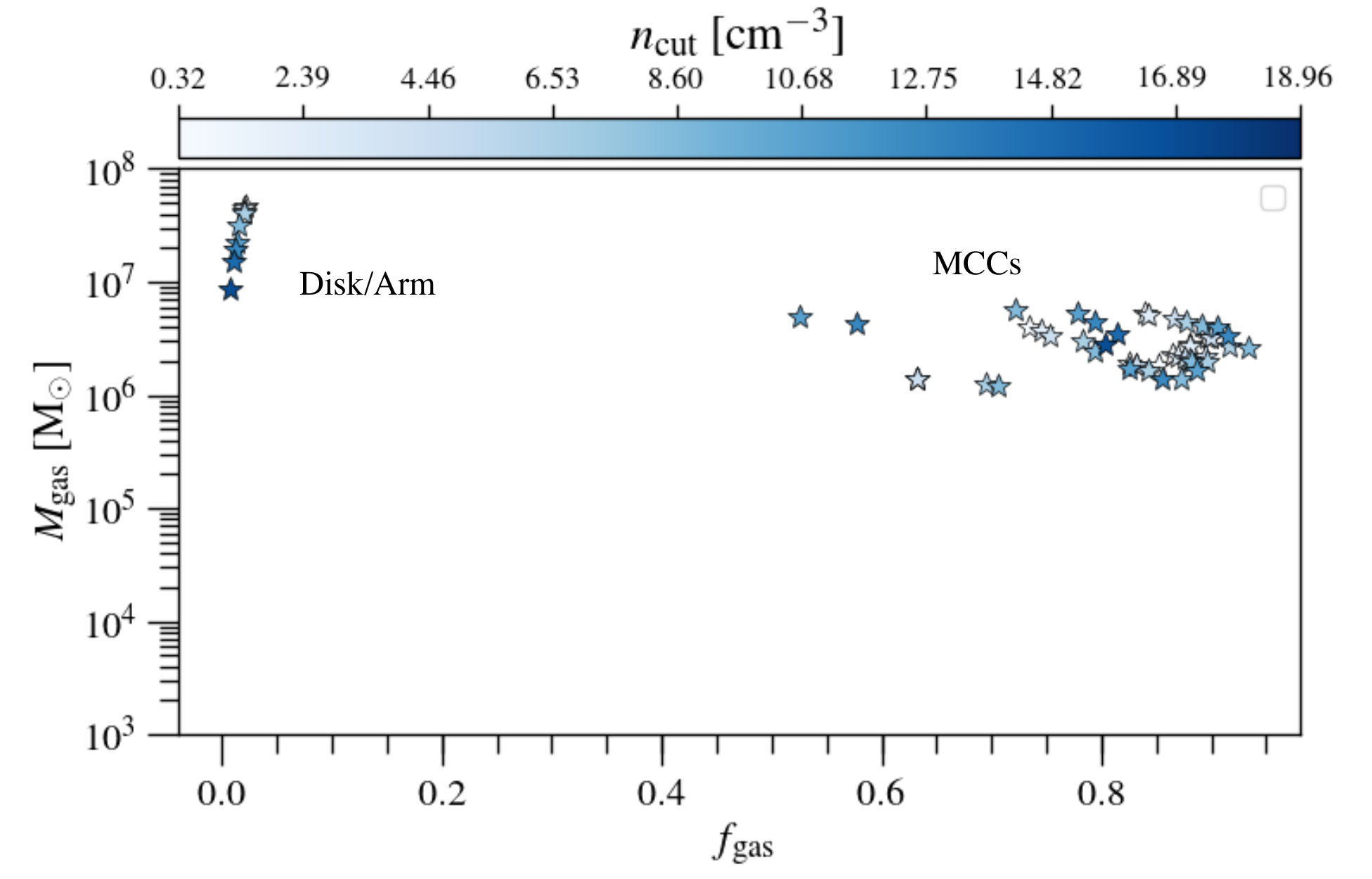}
\caption{Cloud mass and gas mass ratio of MGCs identified in the accreting phase of \flower using different $n_{\rm cut}$ (see colorbar). The most massive structures identified with a low $f_{\rm gas}$ correspond to the molecular disk/arms of \flower (see also \Fig{vv}).
Thus, these structures encompass a large amount of stellar mass that is already assembled in \flower. These disk/arm structures are excluded in the discussion of MGC dynamics in the remainder of this paper.
\label{fig:stellarRatio16}}
\end{figure}

\begin{figure}
\centering
\includegraphics[trim=0 0 0 0, clip, width=0.5\textwidth]{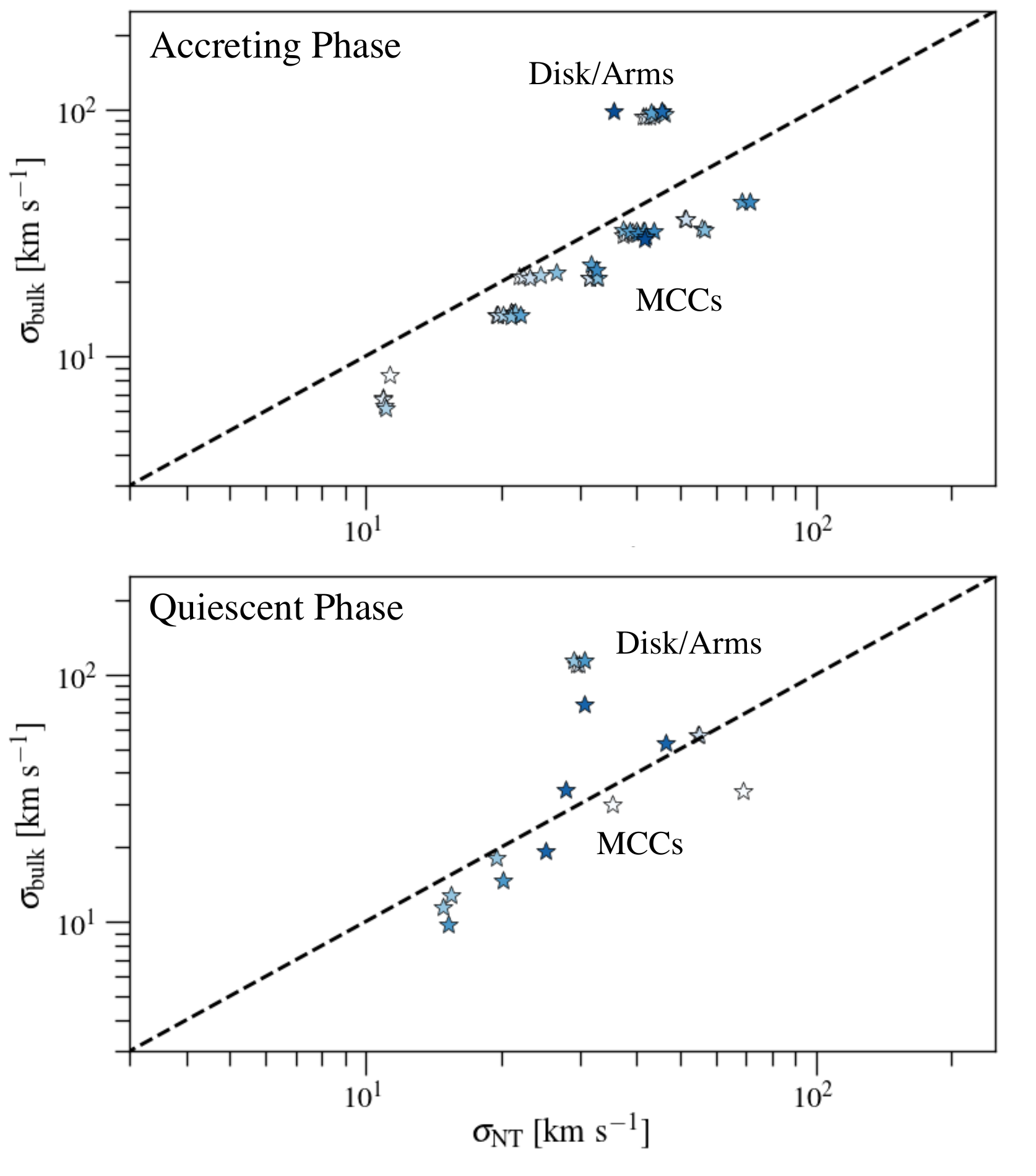}
\caption{Gas velocity dispersions resulting from bulk versus turbulent motions for molecular structures
identified across all values of $n_{\rm cut}$, shown using the same color-coding as \Fig{stellarRatio16}.
Top and bottom panels show structures identified in the accreting phase and the quiescent phase, respectively.
The accreting phase of \flower displays a more disturbed morphology compared to the quiescent phase, which displays a disk-like morphology (see \Fig{MGC}).
The dashed black lines show a one-to-one mapping between the axes to facilitate comparison.
Velocity dispersion from turbulent motions dominates over bulk motions for MGCs in the accreting phase, whereas
$\sigma_{\rm bulk}$ is comparable to $\sigma_{\rm NT}$ for some MGCs in the quiescent disk-like phase.
\label{fig:vv}}
\vspace{0.5em}
\end{figure}

The most massive structures ($M_{\rm gas}>$10$^7$\,\Msun) correspond to the molecular disk of the galaxy (see e.g., \Fig{MGC}), which is identified as a single component at low $n_{\rm cut}$. This main disk component occupies the top left corner of \Fig{stellarRatio16}, with a low gas mass fraction of $f_{\rm gas}< 5\%$. The velocity dispersion of such a structure is dominated by
the bulk motion term\footnote{The bulk velocity dispersion term is likely associated with the rotational velocity, see \citet{Kohandel19a} for kinematical analysis of \flower.}, as shown in \Fig{vv}, where we compare the velocity dispersions of MGCs resulting from bulk ($\sigma_{\rm bulk}$) versus non-thermal turbulent motions ($\sigma_{\rm NT}$) in the accreting and quiescent phases of \flower. The accreting phase of \flower displays a disturbed morphology, whereas the quiescent phase displays a disk-like morphology (\Fig{MGC}).
Excluding the structure corresponding to the molecular disk, the velocity dispersions of most MGCs in the accreting phase of \flower are dominated by turbulent motions, while  in the quiescent disk-like phase, $\sigma_{\rm bulk}$ is at least comparable to $\sigma_{\rm NT}$ for almost half the MGCs.
For these reasons, the most massive MGCs are excluded in the discussion of the dynamics (\Sec{diss}), as they are dominated by large-scale shear; however, we include them in the scaling relation plots (\Fig{larsons_single} and \Fig{alpha16-28}), as would be done for an observational study.

Considering only the MGCs identified at the highest density threshold, the mass of MGCs is $\simeq10^{6.5}$\,\Msun. We note that similarly massive molecular structures (few times 10$^7$\,\Msun\footnote{See \S{3.1} of \citet{Behrendt16a}.}) have been reported in idealized closed-box isolated galaxy simulations done at higher resolution (e.g., a maximum resolution of 3\,pc in the 48\,kpc box studied by \citealt{Behrendt16a}). However, the stellar feedback, IGM, merger and accretion histories are not properly modeled in such simulations, e.g. see \citet{grisdale:2017,grisdale:2019} for the importance of feedback in shaping the properties of clouds forming in local simulated galaxies. That said, the similar mass range found in MGCs of \flower\ is reassuring --- our results are not far off in spite of the limited resolution ($l_{\rm cell}\simeq$\,30\,pc).

The local sound speed of all MGCs identified is typically much smaller than their non-thermal (turbulent) velocities,
as non-thermal pressure dominates thermal pressure for dense gas \citep{Pallottini17b}, i.e., $c_s^2 \ll \sigma_{\rm NT}^2$. In particular, the average Mach number for MGCs identified at the highest density threshold is $\bar{\mathcal{M}}\,\simeq6$. This is consistent with the analysis done in \citet{Vallini18a}, that finds a global Mach number of $\mathcal{M} \sim 10$ for \flower.

\subsection{Single Evolutionary Stage}\label{sec:singless}

\begin{figure*}
\centering
\includegraphics[trim=0 0 0 0, clip, width=\textwidth]{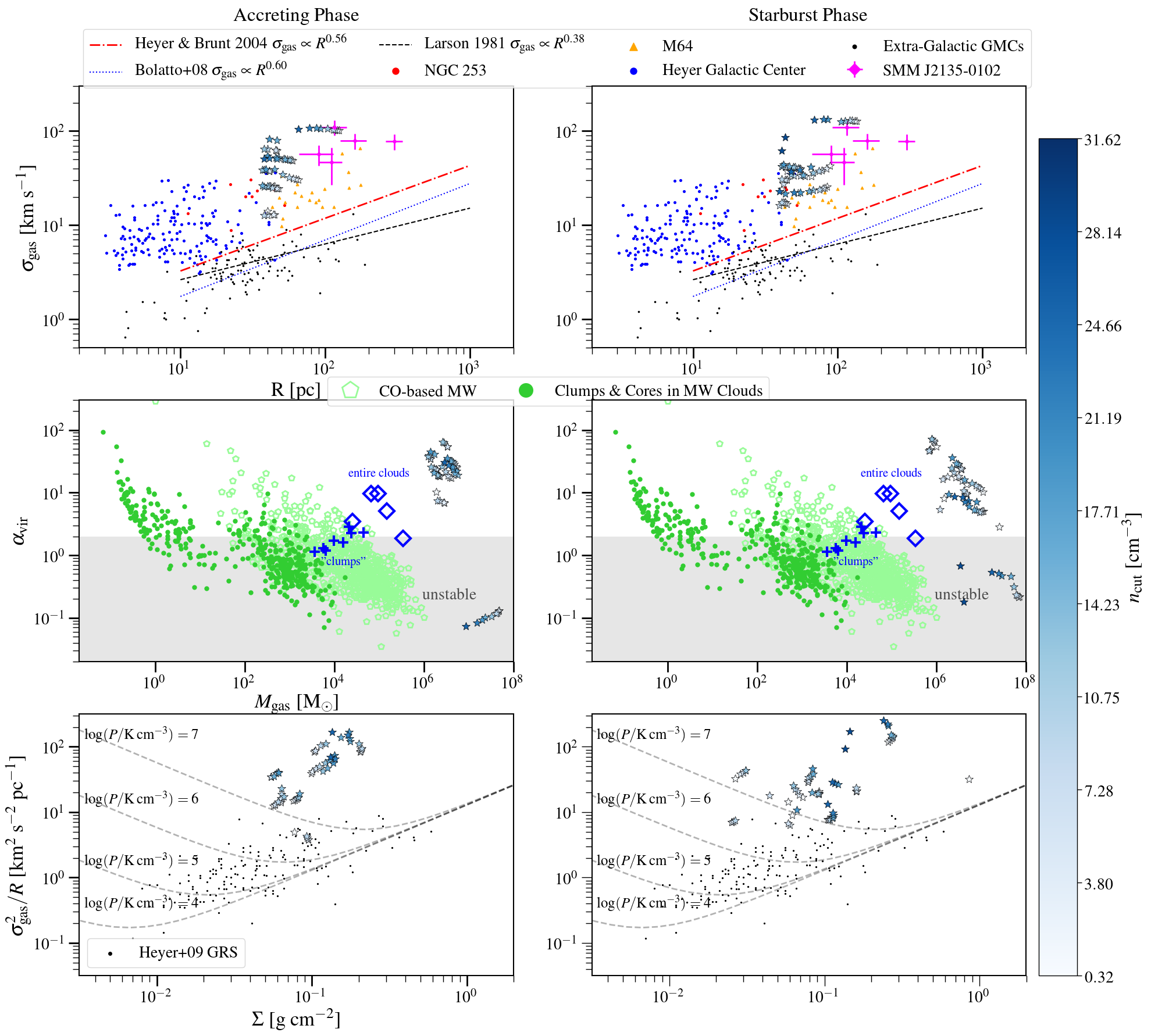}
\caption{
Linewidth-size relation (top), $\alpha_{\rm vir}$-mass relation (middle), and $\sigma_{\rm gas}^2/R$-$\Sigma_{\rm gas}$ relation (bottom) for MGCs (star symbols) identified in the two most extreme evolutionary stages of \flower\ --- accreting (left) and starburst (right).
Star symbols are color-coded by density thresholds $n_{\rm cut}$, as illustrated by the colorbar shown on the right.
Stability of the biggest structures is strongly influenced by the stellar component, given their low $f_{\rm gas}$ (see \Fig{stellarRatio16} and \Sec{dist}).
The gray dotted lines shown in the bottom panels correspond to the various annotated external pressures needed in order for the gas to be in equilibrium, see Equation~\ref{eqn:v0}. Literature data are taken from \citet{Larson81a, Heyer04a, Rosolowsky05a, Bolatto08a,
Swinbank11a, Leroy15a, Kauffmann17a}, and \citet{Kauffmann17b}. MGCs in the starburst phase have lower $\alpha_{\rm vir}$ than in the accreting phase.
\label{fig:larsons_single}}
\end{figure*}

\begin{figure*}
\centering
\includegraphics[trim=0 0 0 0, clip, width=1.05\textwidth]{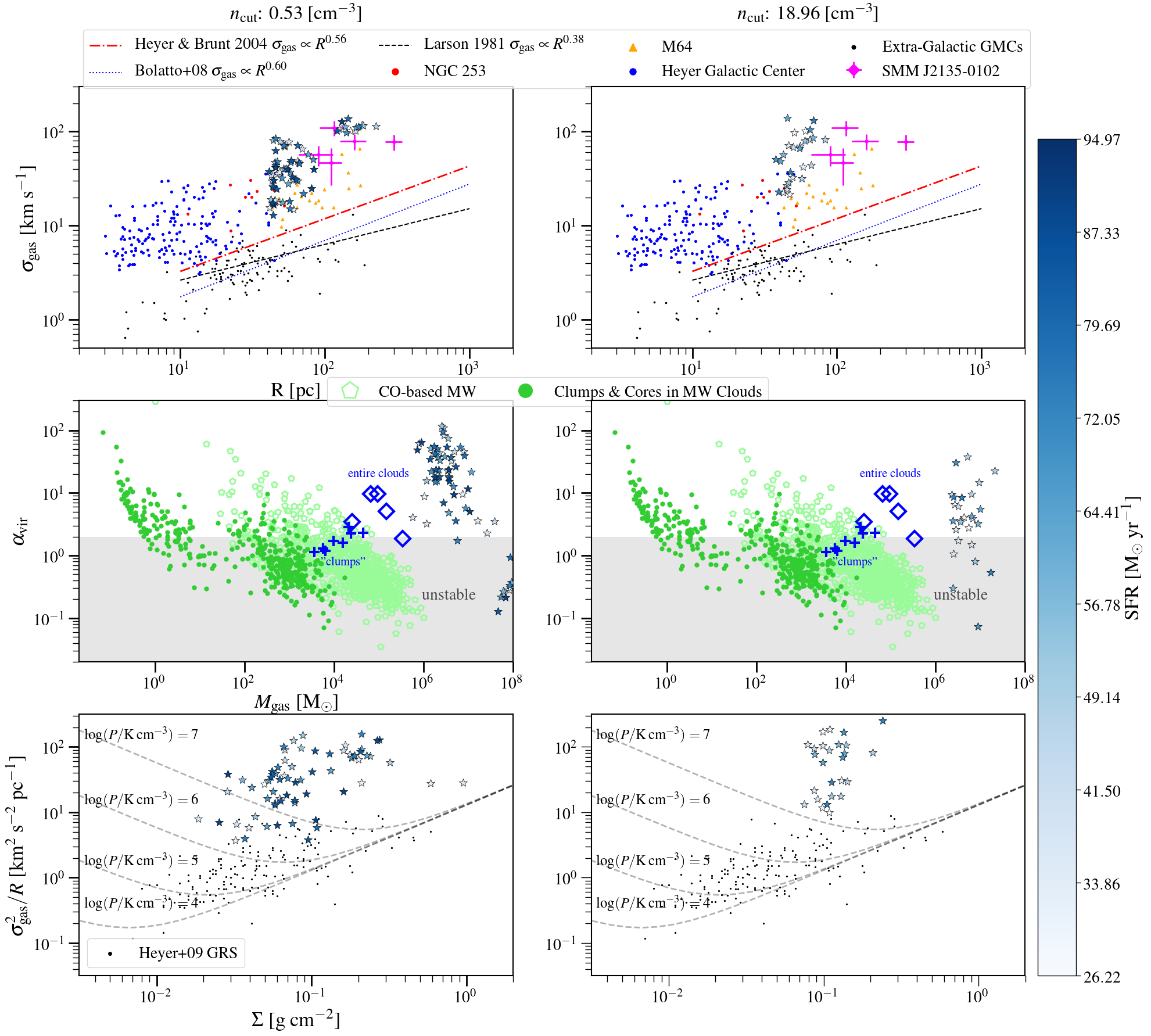}
\caption{Same as \Fig{larsons_single}, except star symbols are showing molecular gas structures identified across all evolutionary stages traced in our simulation, which are color-coded by the SFR of \flower in those stages (see colorbar on the right). Left panels show structures identified using a low density threshold of $n_{\rm cut}$\eq0.53\,\cc and the right panels show those identified using a high density threshold of $n_{\rm cut}$\eq18.96\,\cc.
The biggest MGCs identified at lowest density thresholds, occupying the top right corner of the top left panel, correspond to the molecular disk and arms of \flower, and are broken down into smaller MGCs at higher density thresholds (see top right panel).
Notably less MGCs in the right panels owing to the fact that only few MGCs with neighboring cells reaching high H$_2$ densities of $n_{\rm cut}\simeq$19\,\cc. The high velocity dispersion in \flower is largely driven by non-thermal pressure (see \Fig{vv}).
\label{fig:alpha16-28}}
\end{figure*}

Here, we focus on the MGC properties in the two most extreme evolutionary stages of \flower\ --- the accreting and starburst phases (see \Sec{sfh}). These properties are plotted in \Fig{larsons_single} together with observational results of nearby and $z$\,$\simeq$2 molecular structures for comparison.

During the accreting phase, MGCs in \flower\ are characterized by large velocity dispersions ($\sigma \simeq 100 \mbox{ km s}^{-1}$) and sizes ($R\simeq 100$\,pc). These values are comparable to those found in starburst galaxies, such as the nearby gas-rich galaxy M64 and the $z$\ssim2 starbursting disk galaxy SMM\,J2135-0102 \citep{Rosolowsky05a, Swinbank11a}, but are higher than those found in the Milky Way and extragalactic GMCs by an order of magnitude \citep{Heyer04a, Bolatto08a}.

While the virial parameter defined as Equation~\ref{eqn:alpha} is typically used in observational studies, as they are designed to probe structures that are composed of mainly molecular gas only, we calculate the virial parameter of MGCs including the influence from the stellar component (i.e., $\alpha_{\rm vir, tot}$ via Equation~\ref{eqn:alpha_tot}) given the high stellar-to-gas mass ratios of some molecular structures identified (see \Fig{stellarRatio16}).
The biggest MGC identified at low $n_{\rm cut}$ and with the highest stellar-to-gas mass ratio ($\gtrsim$\,1; \Fig{stellarRatio16}) corresponds to the central main disk of \flower (see top panels of \Fig{MGC} and \Sec{distribution}). The large amount of stellar mass already assembled in \flower explains the low {\em total} virial parameter seen in the middle left panel of \Fig{larsons_single}, where the stellar gravitational potential influences the overall stability. On the other hand, the MGCs in this phase are mostly stable\footnote{We check that the gas-only virial parameter of MGCs is comparable to the {\em total} virial parameter.}, with virial parameter  within the range observed in the molecular structures
of the Milky Way but higher than its average value, though the MGCs are more massive.
The higher than typical MW value is linked to the higher turbulence of these systems.

Motivated by observational studies, we also plot the $\sigma_{\rm gas}^2/R$ ratio and gas surface density of MGCs in the accreting phase of \flower in the bottom left panels of \Fig{larsons_single}. In the same plot, we show those observed in the Galactic Ring Survey (GRS) of the Milky Way \citep{Heyer09a} for comparison. Dashed lines in the figure show the loci along which the annotated external pressures are needed for any molecular clouds in equilibrium to have certain linewidths for a given set of surface densities (see \Sec{PVE}).
At a given gas surface density, MGCs display a range of $\sigma_{\rm gas}^2/R$.
Such variation in $\sigma_{\rm gas}^2/R$ is observed in GMCs in the central and outer part of the Milky Way \citep{Oka01a, Heyer09a}.

If constant column density (Larson's third relation) were truly a
fundamental property of molecular clouds,
and if virial equilibrium were a universal property of molecular clouds with negligible surface pressure, one would expect a single point in the $\sigma_{\rm gas}^2/R$--$\Sigma$ relation at that
mass surface density $\Sigma$.
The variation in $\sigma_{\rm gas}^2/R$ with $\Sigma$ seen here suggests that column density is not a fundamental property of
molecular clouds in agreement with \citet{Heyer09a}.
Previous studies suggest that the observed mass-size relation of molecular clouds
is solely a result of $\rho\propto M R^{-3}$ and may be an artifact of the limited range of
column densities a specific molecular line tracer is sensitive to \citep[see e.g.,][]{Ballesteros02a, Ballesteros11a}.

In the starburst phase, MGCs have velocity dispersion and sizes similar to those in the accreting phase; however, MGCs in the starburst phase span a wider range in gas surface density (bottom panels of \Fig{larsons_single}) and have lower virial parameters.
That is, MGCs in the starburst phase are more susceptible to collapse, as expected. The gas surface density of MGCs in both accreting and starburst phases is higher than that in the solar neighborhood of the Milky Way and the Large Magellanic Cloud (LMC), but comparable to those observed in starburst galaxies and the nearby ultra-luminous IR galaxies \citep[ULIRGS;][]{Boulares90a, Scoville91a, Weiss01a, Hughes10a, Leroy15a}.
%

\subsection{Adopted Density Threshold Dependence}\label{sec:ncut}

As seen in \Sec{dist}, the cuts used to identify the molecular structures are based on the H$_2$ number density, which ranges from
n$_{\rm H2}$\ssim0.1\,--\,30\,\cc. We do not explicitly use a minimum H$_2$ fraction to select the structures.
Among the various cuts, two physical regimes are identified, corresponding to materials in the disk/interarm material and MGCs (see \Fig{stellarRatio16}).
The former have n$_{\rm gas}<100$\,\cc and n$_{\rm H2} < 5$\,\cc, whereas the latter have
n$_{\rm gas}$\ssim$100$\,\cc and n$_{\rm H2} \gtrsim10$\,\cc (see also Figs. 5 and 6 of \citet{Pallottini17b}
for details\footnote{In particular, their analysis of the Minkowski function of the molecular density field (see Fig. 6 therein)}).
Considering that $f_{\rm H2}$\eq$\rho_{\rm H2} / \rho_{\rm H,total}$\eq$\rho_{\rm H2}/(\mu \rho_{\rm gas})$\eq
$m_{\rm H2} n_{\rm H2} / (m_p\,\mu\,n_{\rm gas})$\ssim2 $n_{\rm H2}/(0.7~n_{\rm gas})$,
for the disk/interarm region $f_{\rm H2}$\ssim14\%, while for the MGCs $f_{\rm H2}$\ssim28\%.
Note that for the lowest value of the H$_2$ cuts, these fractions might be lower; however,
removing regions selected with such cuts yields no qualitative change to the physical interpretation resulting from the analysis.

We investigate possible variations in the dynamics of the molecular structures of \flower by adopting different values of $n_{\rm cut}$ to test the robustness of our results against the choice of density threshold.
That is, how sensitively are the structure properties, and thus, the results presented in \Sec{singless}, dependent on the choice of density thresholds?
Observationally, this effect would be mimicked by e.g., adopting different molecular gas tracers since they have different critical densities.

The sizes of MGCs are dependent on the choice of $n_{\rm cut}$ in the following ways. As mentioned in \Sec{dist}, the most massive cloud identified at low H$_2$ gas density thresholds corresponds to the molecular disk of \flower, which is broken down into multiple smaller MGCs at higher $n_{\rm cut}$\footnote{That said, there are fewer MGCs at highest $n_{\rm cut}$ than the lowest $n_{\rm cut}$ since there are fewer cells in the simulations with correspondingly high H$_2$ gas densities.}.
Second, excluding such structure, some MGCs are further broken down into smaller MGCs at higher $n_{\rm cut}$ (see top panels of \Fig{alpha16-28}). At the resolution limit of our simulation, most MGCs studied in this work have $R\simeq$\,50\,pc (see also \Fig{dist}).
The gas velocity dispersion $\sigma_{\rm gas}$ of MGCs, on the other hand, is rather insensitive to the actual value of $n_{\rm cut}$. This lack of variation is reassuring: our inference on the velocity dispersion of \z$\sim$\,6 MGCs in relation to those observed in nearby and \z$\sim$2 galaxies is not biased by our choice of $n_{\rm cut}$ in this work.

At the highest density cut, only the densest gas structures are identified\footnote{In the simulation, the
densest molecular regions reaches gas density of $n_{\rm gas}$\ssim10$^3$\,\cc, which is comparable to the critical density
of low-$J$ CO lines (see e.g., \citealt{Vallini18a}); however, we emphasize that we do not resolve the internal structure of molecular clouds
at the resolution of the simulation.}. The virial parameter of these MGCs are, on average, lower than those identified at low $n_{\rm cut}$, and are thus, more unstable against collapse (see middle right panel of \Fig{alpha16-28}).

\section{Discussion}\label{sec:diss}

\subsection{Synthetic vs Observed MGC Properties} \label{sec:diss1}

In this section, we compare the observable quantities of MGCs obtained from the simulations (e.g., $\sigma_{\rm gas}$, $R$, $M_{\rm gas}$, $\Sigma_{\rm gas}$) with those of molecular gas structures from \obs at lower redshift.

The velocity dispersion of MGCs in \flower is similar to those observed in $z$\ssim2 spatially resolved studies of gas-rich star-forming galaxies,
spanning a range of $\sigma_{\rm gas}\simeq$\,20$-$80\,\kms (\Fig{larsons_single}; see e.g., \citealt{Swinbank11a}).
However, the sizes of MGCs are almost a factor of two smaller. This suggests that the sizes of \highz molecular gas structures reported in the literature
are probably limited by the spatial resolution.
The velocity dispersions of some MGCs are also comparable to the most turbulent gas clouds observed in the inner Milky way and nearby gas-rich galaxies (e.g., M64; \citealt{Oka01a, Rosolowsky05a, Heyer09a, Leroy15a}), which lie along the locus of $\sigma_{\rm gas}\propto R^{0.56}$. Such high velocity dispersion in \flower is dominated by non-thermal energy (\Fig{vv}) and results from the injection of kinetic energy from recent \SF as \flower is assembling its stellar mass (given the high Mach numbers of MGCs; \Sec{distribution}). In fact, by the accreting stage (see \Fig{SFH}a) at $z\simeq$\,7.2, \flower has assembled a stellar mass of $M_\star$\eq7.5\E{9}\,\Msun.
This contrasts with \citet{Krumholz18a}, who show that numerical models and analytic arguments point to feedback generally not being able to produce velocity dispersions of more than 10--20 km~s$^{-1}$, with higher velocities likely being produced by Toomre instability driven radial inflow.

The total pressure\footnote{Non-thermal pressure dominates the total pressure in these high density regions.} of MGCs identified at the highest $n_{\rm cut}$ ranges from $P/k_B\simeq10^{7-9}$\,{\rm K}\,{\rm cm}$^{-3}$, with a median of $10^{7.6}$\,\,{\rm K}\,{\rm cm}$^{-3}$. Such high pressures are comparable to those observed in local ULIRGs; however, the molecular clouds in these local galaxies are concentrated within their central regions \citep{Downes98a, Sakamoto08a}. In our simulated galaxy, the high pressure MGCs are found throughout the disk. This difference likely stems from the different physical mechanisms giving rise to the highly turbulent nature of these molecular structures.
While MGCs in both local ULIRGs and \flower may form via gravitational instability (see e.g., \citealt{McKee07a}), MGCs in the local merger-driven ULIRGs
are likely formed by shock compression and cloud-cloud collisions that funnel large amounts of gas from the progenitor galaxies towards the central region (\citealt{Tan00a, Wu18a}; see \Sec{Qeff} for Toomre-$Q$ analysis on \flower). As shown
by \citet{Kohandel19a} and \citet{Gallerani18a}, the highly turbulent nature seen in the MGCs of \flower result from extra-planar flows and higher velocity accretion/SN-driven outflows. The presence of extra-planar flows may also be the dominant mode for forming the highly supersonic massive MGCs observed in gas-rich star-forming galaxies at $z$\ssim2. For instance, \citet{Swinbank11a} report ISM pressure of $P/k_B\sim$10$^7$\,K\,\cc and molecular gas mass of $M_{\rm gas}\simeq10^{8-9}$\,\Msun in the star-forming regions of a gas-rich starburst galaxy at \z\eq2.3 based on spatially resolved CO line \obs.

We do not find any major quantitative differences (see \Fig{larsons_single}) in the MGCs of \flower with respect to those observed in the nearby Universe between its accreting and starburst phase, which are separated by $\simeq$300\,Myr. However, MGCs in the \SB phase have lower $\alpha_{\rm vir}$ compared to the accreting phase. This is expected for \SF to proceed.

\subsection{Toomre Parameter and Stability of MGCs} \label{sec:Qeff}

\begin{figure*}
\centering
\includegraphics[width=0.9\textwidth]{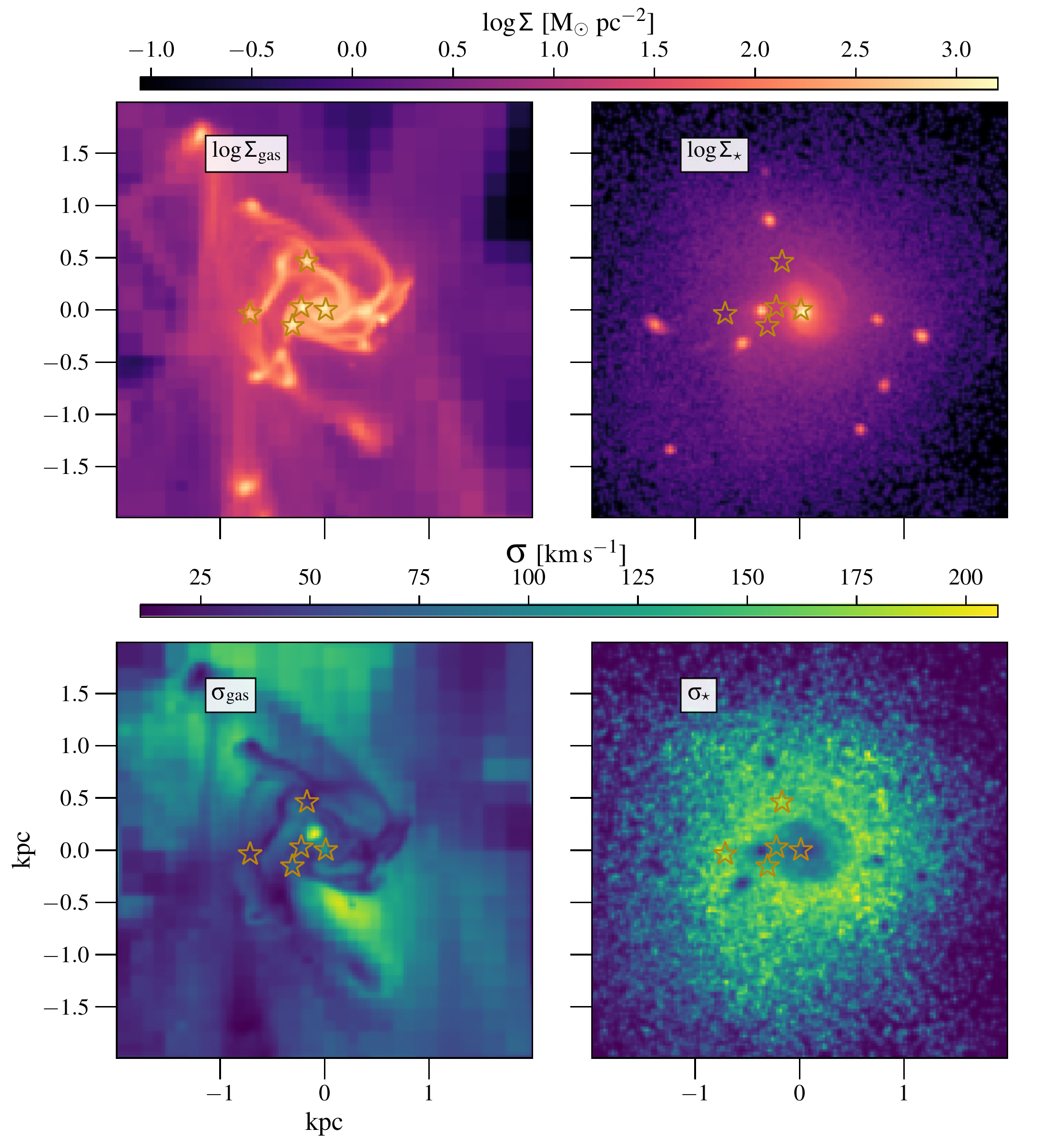}
\caption{
Surface density maps of the gas (top left) and stellar (top right) components of \flower (accreting phase) and their radial velocity dispersion maps projected onto the $xy$-plane (bottom panels). Center of mass positions of MGCs within $\sim$1.5\,kpc of \flower identified with $n_{\rm cut}$\eq6.81\,\cc are overplotted as star symbols as an illustrative example.
\label{fig:sigma}}
\end{figure*}

\begin{figure*}
\centering
\includegraphics[trim=0 0 0 0, clip, width=\textwidth]{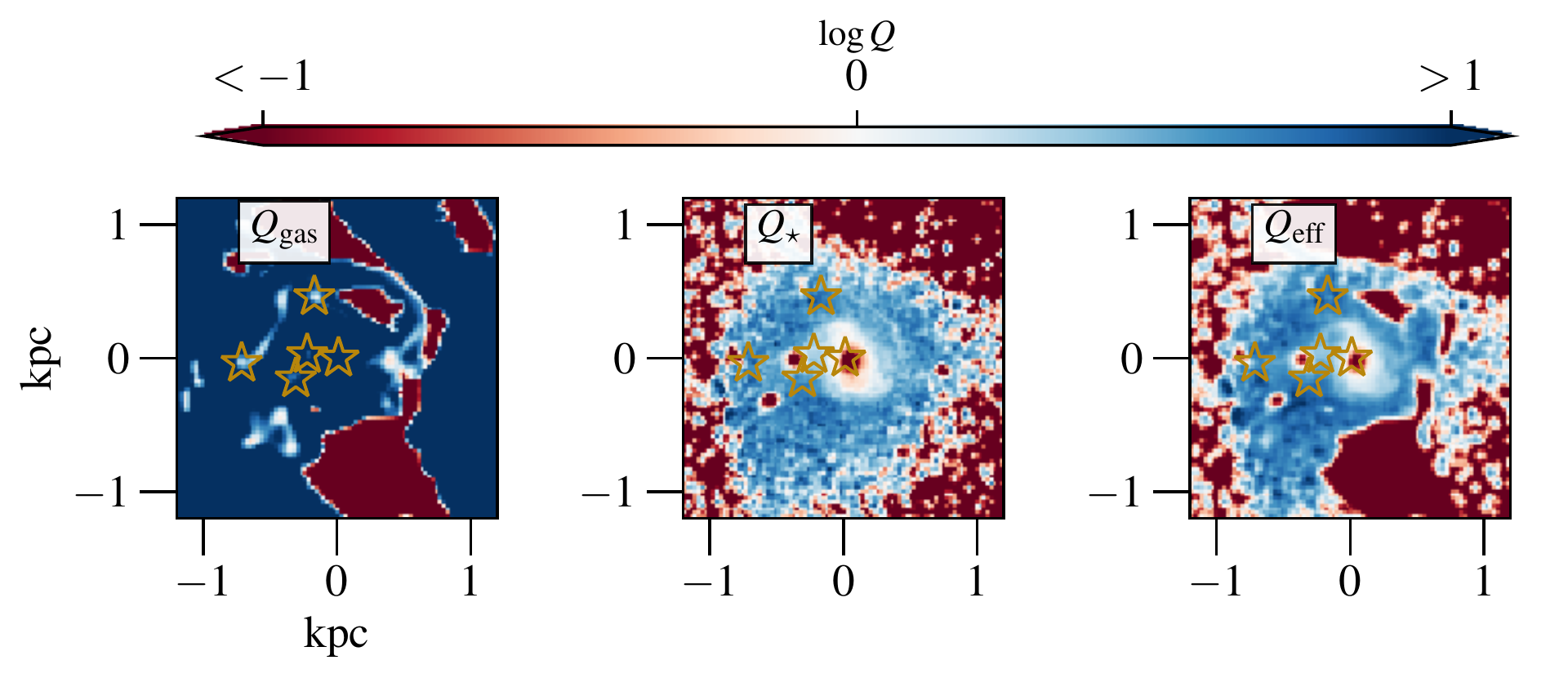}
\caption{
Toomre $Q$ maps derived from the {\bf central region} of \flower. Contribution from the gas ($Q_{\rm gas}$) and the stellar ($Q_\star$) components is shown in the left and middle panels, respectively. The effective two-component Toomre $Q_{\rm eff}$ parameter map is shown in the right panel.
All maps are projected onto the plane of the disk, and a gaussian smoothing of width 30\,pc has been applied to the maps.
Positions of MGCs identified with $n_{\rm cut}$\eq6.81\,cm$^{-3}$ are overplotted as star symbols. Some MGCs lie in regions of $\log{Q_{\rm eff}}\gtrsim0$, where regions of $\log{Q_{\rm eff}}\lesssim0$ are likely gravitationally unstable.
The close resemblance of the $Q_\star$ and $Q_{\rm eff}$ maps in the {\em central region} reflects the fact that the stellar component plays an important role in governing the stability of MGCs against $m=0$  perturbations in {\bf this} relatively evolved and enriched system at high redshifts.
\label{fig:Qeff}}
\end{figure*}

We define Toomre $Q$ parameters for the gas, stars, and the effective two-component $Q_{\rm eff}$ in \Sec{Q}. Maps of the corresponding Toomre $Q$ are shown in \Fig{Qeff} (see also \Fig{h2density} for example of MGCs identified in this evolutionary stage of \flower using different $n_{\rm cut}$).

Close resemblance of the $Q_{\rm eff}$ and $Q_{\star}$ maps indicates that contributions from the stellar component play an important role in governing the stability of the MGCs against perturbations. This can be understood since stars in \flower dominate the central part of the galaxy in mass, so their gravitational potential provides a non-negligible contribution to the instability.
Similarly, contribution from the thickness of the disk is important in \flower since its disk is warped and has a scale height-to-radius ratio of $h/r_{\rm gal}$\ssim150\,pc/1\,kpc\,$\simeq$\,0.15.
That is, some MGCs are found in regions of $Q_{\rm gas}$\ssim1, which is consistent with the expectation that they correspond to regions of high surface densities that are gravitationally unstable. Note, however, that when including the stabilizing effects due to the stellar potential of \flower and the thickness of its disk (via $\sigma_z$, i.e., $Q_{\rm eff}$; see Equation~\ref{eqn:q_eff}), some of these MGCs are consistent with $Q_{\rm eff} > 1$.
This demonstrates the importance of accounting for stellar contribution and disk thickness when examining the stability of molecular gas structures. This consideration is especially relevant for the relatively evolved and enriched systems at high redshift that are preferentially being imaged at high resolution with ALMA. In the outer regions of \flower, on the other hand, $Q_{\rm eff}$ resembles $Q_{\rm gas}$, with both $\log{Q_{\rm gas}}$ and $\log{Q_{\rm eff}} < -1$. Notably, these are the more gas-rich regions (see \Fig{sigma}).

The large virial parameters ($\alpha_{\rm vir}>$\,2) seen in some MGCs of \flower can be understood by first noting that fragmentation {\em can} happen in regions of low $Q$ (if we only consider instability against axisymmetric perturbations), but further evolution and gas collapse depends on the equation of state of the gas.
Such fragmentation is expected to take place at the critical scale length $\lambda_{\rm crit} < 2 \pi^2 G \Sigma / \kappa^2$. Second, this fragmentation scale is greater than the typical size of GMCs.
This could be interpreted as a result of instability setting the scales for fragmentation, {\em but the truly star-forming regions correspond to the collapsing, denser, and cooler molecular structures that are on smaller scales}.
Thus, the high virial parameter found for some MGCs in \flower indicates that they are not the collapsing structures and are found in regions of $Q_{\rm eff}>1$. On the other hand, MGCs in (the denser) regions with a lower Toomre $Q_{\rm eff}$ parameter are more unstable, and \SF may take place within its star-forming {\it clumps} and {\it cores} on smaller scales, where energy quickly dissipates.

\section{Summary and Conclusions}      \label{sec:conclusion}

Properties of the star-forming ISM of galaxies near the Epoch of Reionization are now within reach with ALMA. While it is possible to obtain sensitive and high fidelity imaging that reveals their gas and dust morphology on sub-kpc scales and even smaller, such \obs remain challenging; logistically, data from multiple cycles pushing to increasing resolution and sensitivity are needed. As such, observational studies of MGCs on such scales are still missing in the literature.
In this work, we aim to understand the origin and dynamical properties of MGCs in prototypical galaxies at the EoR in numerical simulations to provide a framework within which upcoming \obs can be compared against to aid in the interpretation.

We study the dynamics of MGCs and their temporal evolution in \flower, a \z$\sim$\,6 prototypical galaxy obtained from the state-of-the-art cosmological zoom-in simulations \ncode{Serra}, which include a chemical network to determine the formation of molecular hydrogen, heating and cooling of the ISM by UV radiation and metal lines, and detailed stellar feedback.
We use a three-dimensional clump-finding algorithm to identify MGCs. We decompose the molecular structures into non-overlapping objects by using a set of H$_2$ density contours ($n_{\rm cut}$) at multiple evolutionary stages, in particular focusing on the accreting phase, starburst phase, and quiescent phase. We extract properties such as mass, size, Mach number, velocity dispersion, gas surface density, and virial parameter ($M_{\rm gas}, R, \mathcal{M}, \sigma_{\rm gas}, \Sigma_{\rm gas}, \alpha_{\rm vir}$) for each MGC and perform a Toomre-$Q$ stability analysis on \flower.

Excluding the main structure (disk of \flower) identified, the typical mass and size of MGC is $M_{\rm gas}\simeq\,10^{6.5}$\,\Msun and $R\simeq\,50$\,pc, respectively. Similarly massive molecular structures have been observed in nearby star-forming and starburst galaxies \citep[e.g.,][]{Keto05a, DonovanMeyer13a, Colombo14a, Leroy15a}, and reported in idealized (no stellar feedback) isolated (no cosmological initial conditions) galaxy simulations done at higher resolution (e.g., $l_{\rm cell}\simeq3$\,pc in the 48\,kpc box studied by \citealt{Behrendt16a}). That said, the similar mass range found in MGCs of \flower\ is reassuring --- our results are not far off despite the resolution limit ($l_{\rm cell}\simeq$\,30\,pc). On the other hand, the gas velocity dispersion
of $\sigma_{\rm gas}\simeq$\,20-10\,\kms is rather insensitive to the adopted density threshold.

Velocity dispersion and gas surface density of MGCs of \flower are systematically higher than Milky Way clouds regardless of the density threshold $n_{\rm cut}$ adopted. These MGCs are, in fact, highly supersonic, with high Mach number of $\bar{\mathcal{M}}\simeq6$. Their velocity dispersions are comparable to those observed in $z$\ssim2 starburst galaxies. A comparison between the bulk and non-thermal velocity dispersions of MGCs indicates that MGCs
are supported by turbulence motions.

High pressure MGCs are found throughout the disk of \flower, with a median pressure of $\bar{P}\simeq10^{7.6}$\,K\,\cc. This is in contrast to the local ULIRGs, where such high pressure has only been observed in molecular clouds concentrated in their central regions \citep{Downes98a, Sakamoto08a}. In ULIRGs, MGCs are formed by shock compression and cloud-cloud collisions that funnel large amounts of gas from progenitor galaxies towards the central region (\citealt{Tan00a, Wu18a}). On the other hand, the highly turbulent MGCs of \flower results from extra-planar flows and high velocity accretion/SN-driven outflows. The presence of such extra-planar flows may also be the dominant mode for forming the highly supersonic massive MGCs observed in gas-rich star-forming galaxies at \z\ssim2, for instance, those reported in \citet{Swinbank11a}.
The variation in $\sigma_{\rm gas}^2/R$ with $\Sigma$ seen in this work is in line with previous studies, suggesting that ``Larson's third relation'' is an artifact arising from observational bias artifact of the limited range of  column densities a specific molecular line tracer is sensitive to \citep[][]{Ballesteros02a, Ballesteros11a}. This would imply column density is not a fundamental properties of molecular clouds and not all molecular clouds are virialized.

We perform virial analysis, as motivated by \obs, to assess the stability of MGCs. On average, the virial parameter of MGCs in the starburst phase of \flower is lower than in the accreting phase, as expected for \SF. Similarly, the virial parameter of MGCs identified at the highest density thresholds are lower than those identified at lower density thresholds.
Close resemblance of $Q_{\rm eff}$ and $Q_{\rm star}$ maps indicates that contribution from the stellar component plays an important role in governing the stability of the MGCs against axisymmetric perturbations, especially in the central part of \flower. Similarly, stabilizing effect due to the thickness of its disk is also non-negligible. This illustrates the importance of accounting for stellar contribution and disk thickness when examining the stability of molecular gas structures, especially in relatively evolved and enriched systems at high redshift that are preferentially being observed now.
Star formation is expected to take place within its star-forming {\it clumps} and {\it cores} on smaller scales, where energy quickly dissipates. This is consistent with the notion that collapsing structures result from gravitational instability occurring within globally stable structures, which are supported by turbulence and rotation on large scale.

This also implies that \obs with spatial resolution better than $\simeq$40\,pc are needed to examine the truly star-forming structures (cores), and thus, \SF in the first galaxies. Such resolution is in principle within reach using ALMA, for which 0\farcs01 resolution images can be attained using the highest frequency bands. With the Next Generation VLA\footnote{\url{http://library.nrao.edu/public/memos/ngvla/NGVLA\_21.pdf}}, which is approximately ten times more sensitive than ALMA, direct observations of molecular gas clouds in galaxies at the EoR on cloud scales will be possible via the higher-$J$ rotational transitions of CO in the 2030s.
In the meantime, cosmological zoom-in simulations, such as \ncode{Serra}, while inherently limited in galaxy statistics and depend on
the sub-grid models adopted, serve as a useful tool for examining and making predictions on the morphology and dynamics of the molecular ISM of the first galaxies.


\section*{Acknowledgements}
We thank Jens Kauffmann, Thushara Pillai, and Mark Swinbank for sharing their data, Caitlin Casey for useful discussions, and the referee for providing constructive
comments on the manuscript.
TKDL gratefully acknowledges support by the NSF through award SOSPA4-009 from the NRAO and support from the Simons Foundation.
AF acknowledges support from the ERC Advanced Grant INTERSTELLAR
H2020/740120.
M-MML acknowledges partial support by the NSF through award
AST18-15461, and hospitality from the Insitute for Theoretical
Astrophysics at the University of Heidleberg.
This work is based on a project developed at the Kavli Summer Program in Astrophysics (KSPA) held at the Center for Computational Astrophysics of the Flatiron Institute in 2018. The program was co-funded by the Kavli Foundation and the Simons Foundation.
We thank the KSPA Scientific and Local Organizing Committees, and the program founder, Pascale Garaud, for supporting the genesis of this work.
We also thank the Center for Cosmology and Particle Physics at the New York University for their hospitality in hosting us after the steam pipe explosion in NYC during the KSPA.
The Flatiron Institute is funded by the Simons Foundation.
This research has made use of NASA's Astrophysics Data System Bibliographic Services.
\software{Python \citep{VanRossum1991}, Astropy \citep{astropy}, Cython \citep{behnel2010cython}, Matplotlib \citep{Hunter2007}, NumPy \citep{VanDerWalt2011}, \ncode{PyMSES} \citep{Labadens2012}, SciPy \citep{scipyref}, and yt \citep{Smith09a,Turk11a}}

\bibliographystyle{yahapj}
\bibliography{master_cleanup,codes}

\end{document}

%% file: preamble.tex
\usepackage[breaklinks,colorlinks,citecolor=blue,linkcolor=red]{hyperref} 
\usepackage{etoolbox}

\makeatletter

\patchcmd{\NAT@citex}
  {\@citea\NAT@hyper@{%
     \NAT@nmfmt{\NAT@nm}%
     \hyper@natlinkbreak{\NAT@aysep\NAT@spacechar}{\@citeb\@extra@b@citeb}%
     \NAT@date}}
  {\@citea\NAT@nmfmt{\NAT@nm}%
   \NAT@aysep\NAT@spacechar\NAT@hyper@{\NAT@date}}{}{}

\patchcmd{\NAT@citex}
  {\@citea\NAT@hyper@{%
     \NAT@nmfmt{\NAT@nm}%
     \hyper@natlinkbreak{\NAT@spacechar\NAT@@open\if*#1*\else#1\NAT@spacechar\fi}%
       {\@citeb\@extra@b@citeb}%
     \NAT@date}}
  {\@citea\NAT@nmfmt{\NAT@nm}%
   \NAT@spacechar\NAT@@open\if*#1*\else#1\NAT@spacechar\fi\NAT@hyper@{\NAT@date}}
  {}{}

\makeatother
\usepackage{tabu}
\usepackage{booktabs}
\usepackage{xspace}
\usepackage{apjfonts}
\usepackage{amssymb, amsmath} 
\usepackage{natbibspacing, natbib}
\usepackage{aas_macros} 
\usepackage{graphics,graphicx}
\newcommand{\pmOne}{\mbox{$^{-1}$}\xspace}

\newcommand{\cii}{[C{\scriptsize II}]\xspace}
\newcommand{\nii}{[N{\scriptsize II}]\xspace}
\newcommand{\oiii}{[O{\scriptsize III}]\xspace}

\newcommand{\ci}{C$\,${\sc i}}

\def\cc{${\rm cm}^{-3}$\xspace}

\newcommand{\Msun}{\mbox{$M_{\odot}$}\xspace}

\newcommand{\kms}{km\,s$^{-1}$\xspace}

\newcommand{\E}[1]{$\times10^{#1}$}

\newcommand{\eq}{\,=\,}
\newcommand{\ssim}{\,$\sim$\,}

\newcommand{\Fig}[1]{Figure~\ref{fig:#1}}
\newcommand{\Eq}[1]{Equation~\ref{eqn:#1}}

\newcommand{\Sec}[1]{\S\ref{sec:#1}}

\newcommand{\ncode}[1]{{\sc #1}}

\newcommand{\z}{$z$\xspace}
\newcommand{\obs}{observations\xspace}

\newcommand{\highz}{high-$z$\xspace}

\newcommand{\SF}{star formation\xspace}

\newcommand{\SB}{starburst\xspace}

\newcommand{\galpop}{galaxy populations\xspace}

\newcommand{\flower}{Alth{\ae}a\xspace}